\documentclass[a4paper,11pt]{article}

\newcommand{\be}{\begin{eqnarray}}
\newcommand{\ee}{\end{eqnarray}}

\def\hbar#1{\slash\hspace{-2.5mm}#1}
\usepackage{epstopdf}
\usepackage{graphicx,graphics}
\usepackage{dcolumn}
\usepackage{bm}
\usepackage{psfrag}
\usepackage{subfigure}
\begin{document}
{\par\raggedleft \texttt{HIP-2009-01/TH}\par}
{\par\raggedleft \texttt{IPPP/09/01}\par}
{\par\raggedleft \texttt{DCPT/09/02}\par}
{\par\raggedleft \texttt{MAN/HEP/2009/4}\par}
{\par\raggedleft \texttt{NORDITA-2009-14}\par}
\bigskip{}

\begin{center}
{\Large\bf Searching for  the  triplet Higgs sector 
via central exclusive production at the LHC}\\[10pt]
M. Chaichian$^{1}$, P. Hoyer$^{1}$, K. Huitu$^{1}$, V.A.
Khoze$^{2,3}$, and A.D. Pilkington$^{3}$\\[10pt]
{\it $^1$Department of Physics, University of Helsinki, and
Helsinki Institute of Physics, P.O. Box 64,
FIN-00014 University of Helsinki, Finland\\
$^2$Institute for Particle Physics Phenomenology, University
of Durham, Durham, DH1 3LE, UK\\
$^3$School of Physics \& Astronomy, University of Manchester,
Manchester M13 9PL, UK}
\end{center}

\begin{abstract}
We discuss the prospects of searching for the neutral Higgs
bosons of the triplet model in central exclusive production
at the LHC. A detailed Monte Carlo analysis is presented for
six benchmark scenarios for the  Higgs boson, $H_1^{0}$, 
 these cover $m_{H_1^0}=$~120, 150~GeV and 
doublet-triplet mixing of $c_H=$~0.2, 0.5 or 0.8.  
We find that, for appropriate values of $c_H$, an excellent 
Higgs mass measurement is possible for the neutral Higgs in 
the triplet model, and discuss how to distinguish the triplet model
Higgs boson from  the Higgs boson of the Standard Model.
\end{abstract}

\section{Introduction}

It is expected that the nature of electroweak symmetry 
breaking will 
be revealed by the LHC experiments in the near future.
In the Standard Model (SM), the electroweak symmetry is spontaneously
broken by a Higgs doublet, which contains a
neutral scalar field that acquires a vacuum expectation value (VEV).
However, several Higgs multiplets typically occur in extensions to the SM.
In supersymmetric models, at least one additional doublet
is required.
In left-right symmetric models, triplets are added to naturally generate
 a small mass for the neutrinos.
Although the new scalars do not always take part in the electroweak symmetry
breaking, they affect the properties
of the Higgs boson through mixing.

Models with an extended Higgs sector typically contain charged
scalars.
A large number of studies \cite{Georgi:1985nv,dcbtheory}
have previously investigated the
possibility of studying the doubly or
singly charged components of higher representation\footnote{For example
the discovery of the $ZW^{(+-)}H^{(-+)}$ and/or
$W^-W^-H^{++}$ vertices would serve as a direct proof
of the non-standard structure of the Higgs sector (see {\it e.g.}
\cite{Johansen:1987yk}).}.
However the charged scalars
may be considerably heavier than the light
neutral bosons. Therefore,
it would be instructive 
 to study the properties of the light neutral Higgs particles
in order to reveal the manifestation of
new representations \cite{Kundu:1995qb}.

Higgs triplets are an especially attractive possibility
\cite{Accomando:2006ga}.
A tiny neutrino mass may indicate that the mass is being generated
by the seesaw mechanism containing the coupling of neutrinos to the triplet.
In addition, composite Higgs models contain several 
multiplets, including the triplet ones.
Triplets also occur in the little Higgs models - see, for example,
\cite{ArkaniHamed:2002qy} and references therein.

Determining that a new detected state is indeed a Higgs boson and
distinguishing it from the Higgs boson of the SM will be far from  trivial. 
This task will require
a comprehensive programme of precision Higgs measurements.
In particular, it will be of  
utmost importance to determine the spin and
$CP$ properties of a new state and to measure precisely its mass, width
and couplings.
In this work, we suggest that the neutral sector
of the triplet
representation can be studied using the central exclusive 
production (CEP) mechanism (see, for example, \cite{KMRpr})
if forward proton detectors are installed at ATLAS and/or CMS,
(see \cite {Albrow:2008pn}).

The structure of the paper is as follows. In Sec.~\ref{sec:hrep}, we consider the
 general properties of models with higher
representations.
We then concentrate on the triplet 
representation in Sec.~\ref{sec:htriplet}; we choose a benchmark model 
with the electroweak $\rho$-parameter
equal to unity at tree-level, although the 
results are quite 
general. 
In Sec.~\ref{sec:cep}, we introduce the central exclusive production process.
 Finally, in Sec.~\ref{sec:simul}, 
 we present a detailed Monte Carlo analysis of the central
  exclusive production of a neutral Higgs boson in the triplet 
  model for a selection of parameter choices.

\section{Models with general Higgs representations \label{sec:hrep}}

We start with the Standard Model gauge group 
SU(2)$_L\times$U(1)$_Y$ for the electroweak sector.
The masses of the gauge bosons are then obtained from the kinetic 
part of Lagrangian,
\begin{eqnarray}
L_{kin}=\sum_k (D^{\mu}\phi_k)^*(D_{\mu}\phi_k) + 
\frac 12 \sum_i (D^\mu\xi_i)^T
(D_\mu\xi_i),
\end{eqnarray}
where $\phi_k$ are complex representations and $\xi_i$ are real ones.
The covariant derivative is
\begin{eqnarray}
D_\mu= \partial_\mu +igW_\mu^aT^a +\frac Y2 g'B_\mu,
\end{eqnarray}
where $T^a$ is the generator of SU(2) in the appropriate representation  
(with Tr($T^aT^b)=\frac 12 \delta^{ab}$) and $Y$ is the U(1) hypercharge.
Here $W^a$ and $B$ are the SU(2) and U(1) gauge bosons respectively,
and the mixing angle $\theta_W$ of the $Z$ boson and photon is 
obtained by diagonalizing the neutral sector.
The $W$ and $Z$ boson masses are given by
\begin{eqnarray}
m_Z^2=(g^2+g'^2)\sum_i T_{3i}^2v_i^2,\quad 
m_W^2=g^2\sum_i T_{3i}^2v_i^2,
\label{Vmasses}
\end{eqnarray}
where $T_{3i}$ is the isospin third component and $v_i$
is the
VEV of particle $i$.
It is clear from Eq.~(\ref{Vmasses}) that the doublet VEV decreases 
when several representations obtain non-vanishing VEVs.
Furthermore, since the left-handed fermions
are in doublets, the charged fermions can only get their masses 
through the Higgs doublet representation\footnote{For the neutral
fermions this is not the case, since the Majorana masses
can be generated through triplets.}, $m_f=y_f v_{doublet}$, and
 the fermion Yukawa coupling, $y_f$, must increase to produce the
fermion masses.
This, for example, leads to an enhancement in the production 
cross section for Higgs production via gluon fusion, where the 
dominant contribution is due to the top quark loop. 
A further enhancement is present in the branching ratio to 
fermion anti-fermion pairs. The possibility arises, therefore,
of observing a very different prediction to that of the Standard Model.

The higher Higgs representations are severely restricted by the
electroweak $\rho$-parameter.
The $\rho$-parameter in the Standard Model is defined by the ratio
of the gauge boson masses,
\begin{eqnarray}
\rho = \frac{m_W^2}{m_Z^2 \cos^2\theta_W},
\end{eqnarray}
which at tree level is exactly unity in the Standard Model. 
The radiative corrections to the $\rho$-parameter have been studied
in the SM up to three-loop level
\cite{vanderBij:1984aj}~-~\cite{Boughezal:2004ef}\footnote{An explicit 
formula for the one-loop correction $\Delta\rho^{(1)}$ in
$\rho =\Delta\rho^{(1)}+\Delta\rho^{(2)}+\Delta\rho^{(3)}$
is given in \cite{Longhitano:1980iz}, two-loop terms are given in
\cite{van der Bij:1983bw}, and three-loop corrections in
\cite{Boughezal:2004ef}.}.
For $m_H\sim 2m_W$, the correction to the $\rho$-parameter is
$\rho -1 \sim -0.00078$+(4-loop and higher corrections). 
For heavier Higgs masses, the absolute value of the negative 
corrections increase. 
In a model with several scalar representations, whose neutral
component develops a VEV, the $\rho$-parameter is given 
at tree level by \cite{hhg}
\begin{eqnarray}
\rho = \frac{\sum_i r_i \left( T_i (T_i+1) -T_{3i}^2)v_i^2\right)}
{\sum_i 2 T_{3i}^2 v_i^2}.
\label{rho1}
\end{eqnarray}
Here $T_i$ is the weak isospin and $r_i=1/2 (1)$ for real (complex) 
representations (see Eq.~(\ref{reps}) for examples).
Finally, the $\rho$-parameter is experimentally constrained to be 
\cite{Yao:2006px},
\begin{eqnarray}\label{eq:exprho}
\rho -1 = 0.0002\begin{array}{l}+0.0024 \\ -0.0009 \end{array},
\label{rho2}
\end{eqnarray}
where the quoted errors are at 2$\sigma$.
As the loop-corrections to the $\rho$-parameter in the SM are
negative,
it can be argued that the $\rho$-parameter favors either a light Higgs
boson or models that result in positive corrections to $\rho$. 

\section{Higgs bosons in a triplet model \label{sec:htriplet}}

In order to fulfill the experimental constraint on the $\rho$-parameter
 in Eqn.~(\ref{eq:exprho}), the
triplet VEV has to be small.
Using Eqs.~(\ref{rho1}) and (\ref{rho2}) one finds that the upper limit 
for the triplet VEV is a few GeV's assuming that there is  
one triplet in addition to one doublet.
An alternative method to satisfy the experimental constraint at 
tree-level\footnote{At
one-loop level, one has to consider renormalization.  It has been
shown in \cite{Chen:2008jg} that
$\rho \neq 1$ at tree-level is acceptable in a real triplet model 
as the experimental measurement of $\rho$ is satisfied after 
calculating higher order corrections.}
is to have representations which add up to $\rho =1$. 
We discuss this next.

\subsection{A model with $\rho =1$}

We consider the
model developed by Georgi and Machacek \cite{Georgi:1985nv}, and further 
studied in 
\cite{Chanowitz:1985ug,Gunion:1989ci,Gunion:1990dt}
in which additional representations are chosen in such a way that
the tree-level value of $\rho$ remains unity. 
The $\rho$-parameter is
fixed to one 
by choosing one complex scalar doublet
($\phi_{Y=1}$) and two triplets, one real ($\xi_{Y=0}$) and one 
complex ($\chi_{Y=2}$).  
These can be written as
\begin{eqnarray}
\phi=\left( \begin{array}{cc} \phi^{0*} & \phi^+ \\
\phi^- & \phi^0 \end{array}\right),\quad
\chi=\left( \begin{array}{ccc} \chi^0 & \xi^+ & \chi^{++}\\ 
\chi^- & \xi^0 & \chi^{+}\\
\chi^{--} & \xi^- & \chi^{0*} 
\label{reps}
\end{array} \right).
\end{eqnarray}
The following sign conventions are chosen:
$\phi^-=-(\phi^+)^*$, $\chi^{--}=(\chi^{++})^*$, $\chi^-=-(\chi^+)^*$,
$\xi^-=-(\xi^+)^*$, and $\xi^0=(\xi^0)^*$.
The VEVs of the neutral components of the Higgs fields are denoted by
$\langle \chi^0\rangle =\langle \xi^0\rangle=b$ and 
$\langle \phi^0\rangle=a/\sqrt{2}$.
For doublet-triplet mixing, the standard notation is employed:
\begin{eqnarray}
c_H\equiv \frac{a}{\sqrt{a^2+8b^2}},\quad 
s_H\equiv \frac{\sqrt{8}b}{\sqrt{a^2+8b^2}},\quad v^2\equiv a^2+8b^2.
\end{eqnarray}
The most general scalar potential for the model, assuming invariance
under $\chi\rightarrow-\chi$, is
\begin{eqnarray}
V&=&\lambda_1 ({\rm Tr} \phi^\dagger\phi -c_H^2v^2)^2 +\lambda_2 
({\rm Tr} \chi^\dagger\chi -\frac 38 s_H^2v^2)^2 \nonumber\\
&&+\lambda_3 ({\rm Tr} 
\phi^\dagger\phi -c_H^2v^2+ {\rm Tr} \chi^\dagger\chi -\frac 38
s_H^2v^2)^2 \nonumber\\
&&-\lambda_4 \left({\rm Tr} \phi^\dagger\phi{\rm Tr} \chi^\dagger\chi
-2\sum_{ij}{\rm Tr} (\phi^\dagger\tau_i\phi\tau_j){\rm Tr} (\chi^\dagger
t_i\chi t_j)\right) \nonumber\\
&&+
\lambda_5\left( 3{\rm Tr} (\chi^\dagger\chi\chi^\dagger\chi)-({\rm Tr} 
\chi^\dagger\chi)^2\right),
\end{eqnarray}
where $\tau_i/2$ are the SU(2) generators in the doublet representation and
$t_i$ in the triplet representation.

\begin{figure}[t]
\centering
\psfrag{cH}{\small$c_H$}
\psfrag{mlambda}{\small$m/\sqrt{\lambda}$ [GeV]}
\psfrag{mone}{\tiny $H_1^0$}
\psfrag{monep}{\tiny $H_1^{0'}$}
\psfrag{mthree}{\tiny $H_3^0$}
\psfrag{mfive}{\tiny $H_5^0$}
\includegraphics[height=5cm,width=0.5\textwidth]{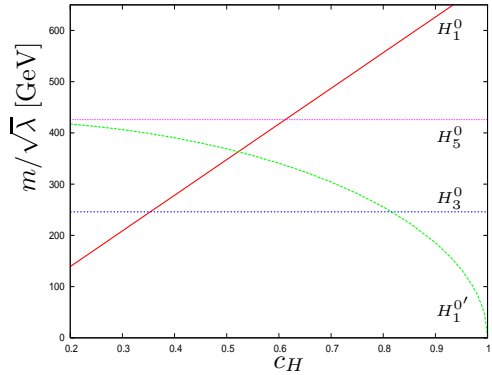}
\caption{\label{fig:mhiggs}
Masses of the Higgs bosons, $m_{H_i}$, divided by the unknown 
$\sqrt{\lambda_i}$ couplings, as a function of $c_H$.}
\end{figure}

As we are interested in this model mainly to illustrate the possibility of  
 studying a neutral triplet Higgs sector, 
it is enough to limit ourselves to the 
case in which $\lambda_3$ is zero and $\lambda_4=\lambda_5$.
The tree-level results of this
triplet model are sufficient for demonstrating the phenomenology of the
higher representations.
In this case, the neutral doublet and triplet do not mix and the
 neutral mass eigenstates are 
\begin{eqnarray}
&&H_1^0=\phi^{0r},\quad H_1^{0'}=\frac
{1}{\sqrt{3}}(\sqrt{2}\chi^{0r}+\xi^0),
\nonumber\\
&&H_3^0=c_H\chi^{0i}+s_H\phi^{0i},\quad
H_5^0=\frac{1}{\sqrt{3}}(\sqrt{2}\xi^0-\chi^{0r}),   
\end{eqnarray}
where $\chi^0=(\chi^{0r}+i\chi^{0i})/\sqrt{2}$.
The masses of the neutral scalars are
\begin{eqnarray}
&&m_{H_1^0}^2=8c_H^2\lambda_1v^2,\quad m_{H_1^{0'}}^2=3s_H^2\lambda_2v^2,
\nonumber\\
&&m_{H_3^0}^2=\lambda_4v^2,\quad 
m_{H_5^0}^2=3(\lambda_5s_H^2+\lambda_4c_H^2)v^2.
\end{eqnarray}
The lightest neutral scalar can be $H_1^0$ if either $c_H$ or $\lambda_1$ is small - 
the $m/\sqrt{\lambda_i}$ values are shown in
Fig.~(\ref{fig:mhiggs}). It should be noted that charged scalars 
exist with the same masses as $H_3^0$ and
$H_5^0$. 

\begin{figure}[tbh]
\psfrag{cH}{\small{$c_H$}}
\psfrag{HVV}{\small{$g_{HVV}/g_{HVV,SM}$}}
\psfrag{YSM}{\small$Y_{Hq\bar q}/Y_{Hq\bar q,SM}$}
\psfrag{H3VV}{\tiny$H_{3}^0VV$}
\psfrag{H5ZZ}{\tiny$H_{5}^0ZZ$}
\psfrag{H5WW}{\tiny$H_{5}^0WW$}
\psfrag{H1VV}{\tiny$H_{1}^0VV$}
\psfrag{H1qq}{\tiny$H_{1}^0q\bar q$}  
\psfrag{H3qq}{\tiny$H_{3}^0q\bar q$}
\mbox{      
	\subfigure[]{\includegraphics[height=5cm,width=.48\textwidth]{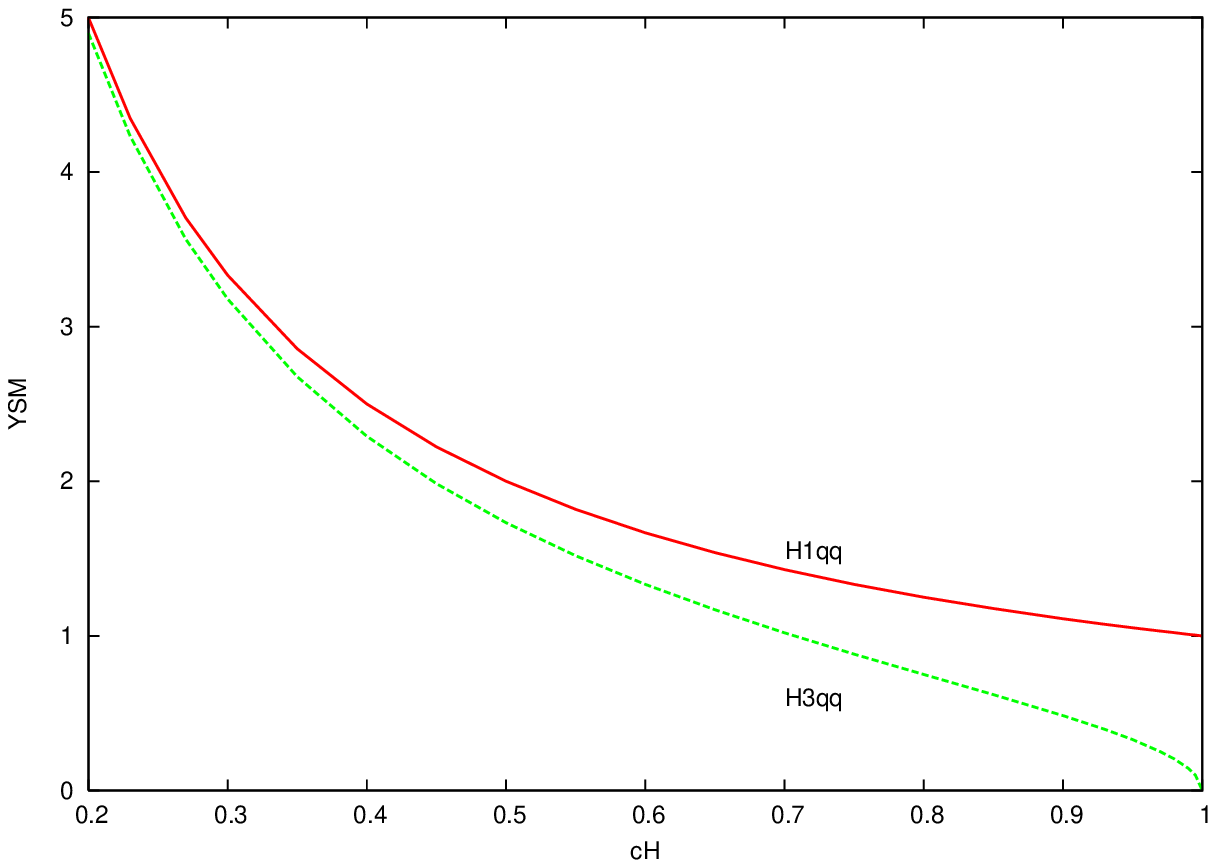}}
	\quad 
	\subfigure[]{\includegraphics[height=5cm,width=.48\textwidth]{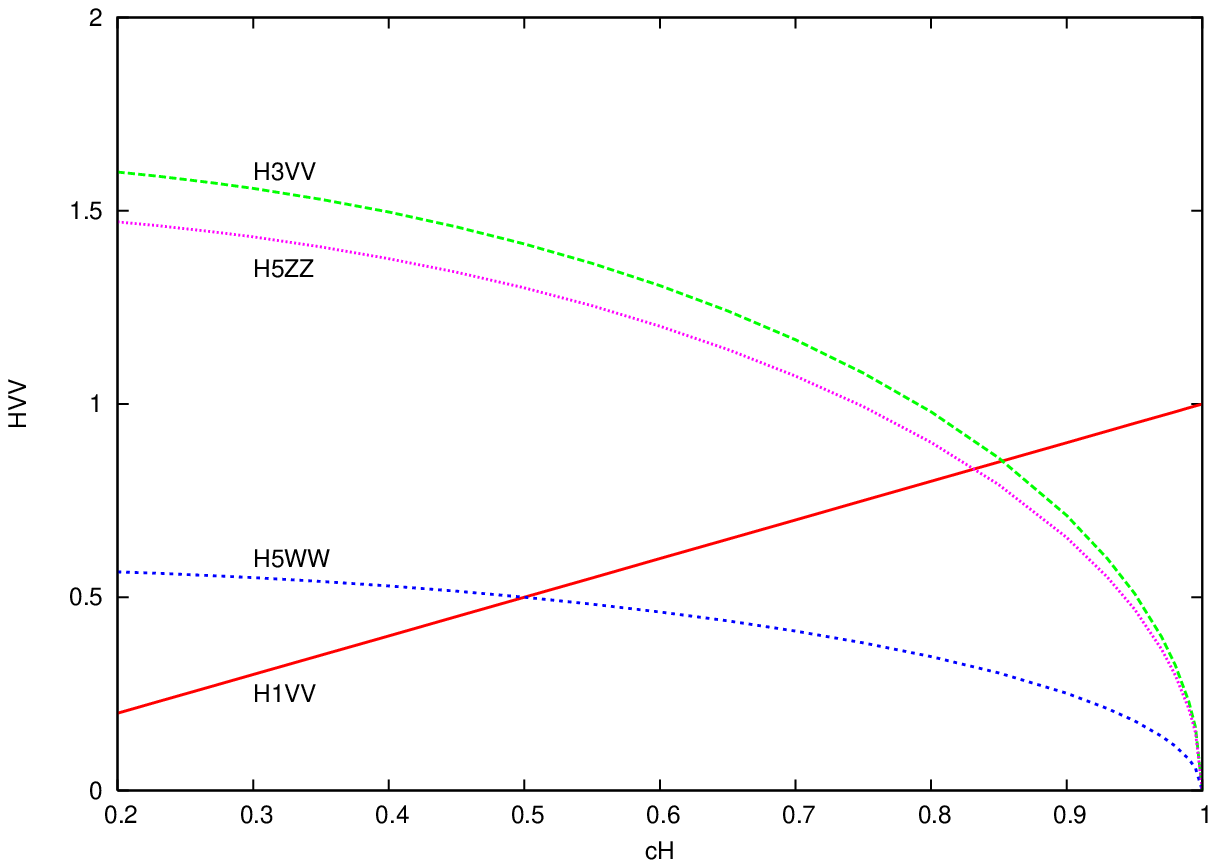}}
 }
\caption{\label{fig:hff}
Couplings of the Higgs bosons to fermions (a) and gauge bosons
(b), normalized by the Standard Model couplings.
}
\end{figure}

The couplings of the neutral scalars to the fermions and the gauge bosons are
\begin{eqnarray}
&&H_1^0 q\bar q: -\frac{gm_q}{2m_Wc_H},\quad
H_3^0 t\bar t: \frac{igm_ts_H}{2m_Wc_H}\gamma_5,\quad
H_3^0 b\bar b: -\frac{igm_bs_H}{2m_Wc_H}\gamma_5,\nonumber\\
&&H_1^0W^+W^-: gm_Wc_H,\quad H_1^0ZZ: \frac {g}{\cos^2\theta_W}m_Wc_H,
\nonumber\\
&&H_1^{0'}W^+W^-: \frac{2\sqrt{2}}{\sqrt{3}}gm_Ws_H,\quad \quad
H_1^{0'}ZZ: \frac {g2\sqrt{2}}{\cos^2\theta_W\sqrt{3}}m_Ws_H,\nonumber\\
&&H_5^{0}W^+W^-: \frac{1}{\sqrt{3}}gm_Ws_H,\quad \quad
H_5^{0}ZZ: -\frac {2g}{\cos^2\theta_W\sqrt{3}}m_Ws_H.
\end{eqnarray}
It is clear that, at tree-level, the coupling of the $H_1^0$ 
to fermions is always enhanced by the factor of $1/c_H$. 
Conversely, the coupling of the $H_3^0$ to fermions
is either enhanced or suppressed, depending on the
ratio of $s_H$ and $c_H$, and the other neutral scalars do not
couple to fermions.
Importantly, the gauge boson couplings to $H_1^0$ are suppressed 
by a factor $c_H$ with respect to the SM and the role of vector boson 
fusion mechanism for $H_1^0$ production is reduced if $c_H$ is small.
The other Higgs couplings to gauge bosons are suppressed by $s_H$.
The $Hq\bar q$ couplings presented in Fig.~\ref{fig:hff} (a)
are considerably enhanced for small $c_H$ in comparison to the SM prediction. 
In Fig.~\ref{fig:hff} (b), the $HVV$ couplings are shown, again  
normalized to the SM couplings.

\subsection{Decays of $H_1^0$ and constraints on the parameters}

In this section, we consider the $H_1^0$ Higgs boson, which becomes the Standard Model Higgs 
boson for vanishing doublet-triplet mixing.
The mass limits for $H_1^0$ can be deduced from the LEP results.
The couplings of the $H_1^0$ to the gauge bosons are smaller than 
in the SM,
leading to reduced production of Higgs bosons in Higgsstrahlung process
\cite{valery}, 
through which the Higgs was expected to be produced at LEP.  
The Higgs boson branching ratio to $b\bar b$ is 74\% for $m_H=114$ GeV
in the Standard Model.
Together with other fermions, the $b\bar b$ decay mode gives the main 
contribution to
the total width of the Higgs boson, and, thus, the Higgs branching ratio 
does not change very much with $c_H$: for $m_H=114$ GeV the 
branching ratio changes to 80\% (81\%) for $c_H=0.5$ ($c_H=0.2$).
If the Higgs boson is lighter, the change is less.
If we assume that the number of $b$-quark pairs gives the Higgs boson
mass limit, it must be heavier than 73 GeV (40 GeV) for $c_H=0.5$ ($c_H=0.2$).
Unitarity further constrains most masses, requiring
them to be less than of the order of 1 TeV \cite{Gunion:1990dt,Aoki:2007ah}.

The Yukawa couplings are constrained by perturbativity, which limits
the $H_1^0$ coupling to top,
\begin{equation}
\frac {g m_{top}}{2m_Wc_H}<\sqrt{4\pi}.
\end{equation}
From this it follows that $c_H> 0.2$, which currently is the most
stringent limit for $c_H$\footnote{Constraints from low energy
precision measurements have been obtained from $Zb\bar b$ vertex,
meson-antimeson mixing and ratios of $b\to u$ to $b\to c$
decays \cite{Haber:1999zh}.  The radiative corrections to the
$Zb\bar b$ vertex give the strongest constraint.
Assuming $m_{H_3}\sim 1$ TeV, one finds $\cos\theta_H > 0.3$ 
with 99.9 \% C.L.  However, since we consider only tree-level
results in this work, we have not used this bound.}.
The latest 95\% confidence limit for the cross section 
$\times$ branching ratio of 
$pp \rightarrow H \rightarrow \tau \tau$ for 
a 120~GeV Higgs boson is observed by the D0 Collaboration 
to be approximately 5~pb given 2.2~fb$^{-1}$ of data \cite{Tevatron}. 
The CDF Collaboration observe a similar limit 
with 1.8~fb$^{-1}$ of data.
 Although this offers no sensitivity to a SM Higgs boson, 
 which has a production cross section of $\sim$1~pb and 
 a branching ratio of $\sim$7\%, one would expect 
increased sensitivity for
the $H_1^0$ in the triplet model for small $c_H$
 (the production cross section is increased by 
 a factor of $1/c_H^2$ w.r.t to the SM as the main production channel 
 would be gluon-gluon fusion via a top quark loop). 
Thus it seems likely that the final combined 
  CDF/D0 results will further constrain the lower $c_H$ limit given 
  the expected integrated luminosity of 8~fb$^{-1}$ per experiment.

\begin{figure}[t]
\psfrag{cH}{\small$c_H$}
\psfrag{BR}{\small$BR$}
\psfrag{bb}{\tiny$b\bar b$}
\psfrag{tautau}{\tiny$\tau^+\tau^-$}
\psfrag{gg}{\tiny$gg$}
\psfrag{gamgam}{\tiny$\gamma\gamma$}
\psfrag{WW}{\tiny$W^+W^-$}
\psfrag{ZZ}{\tiny$ZZ$}
\mbox{      
	\subfigure[]{\includegraphics[height=5cm,width=.48\textwidth]{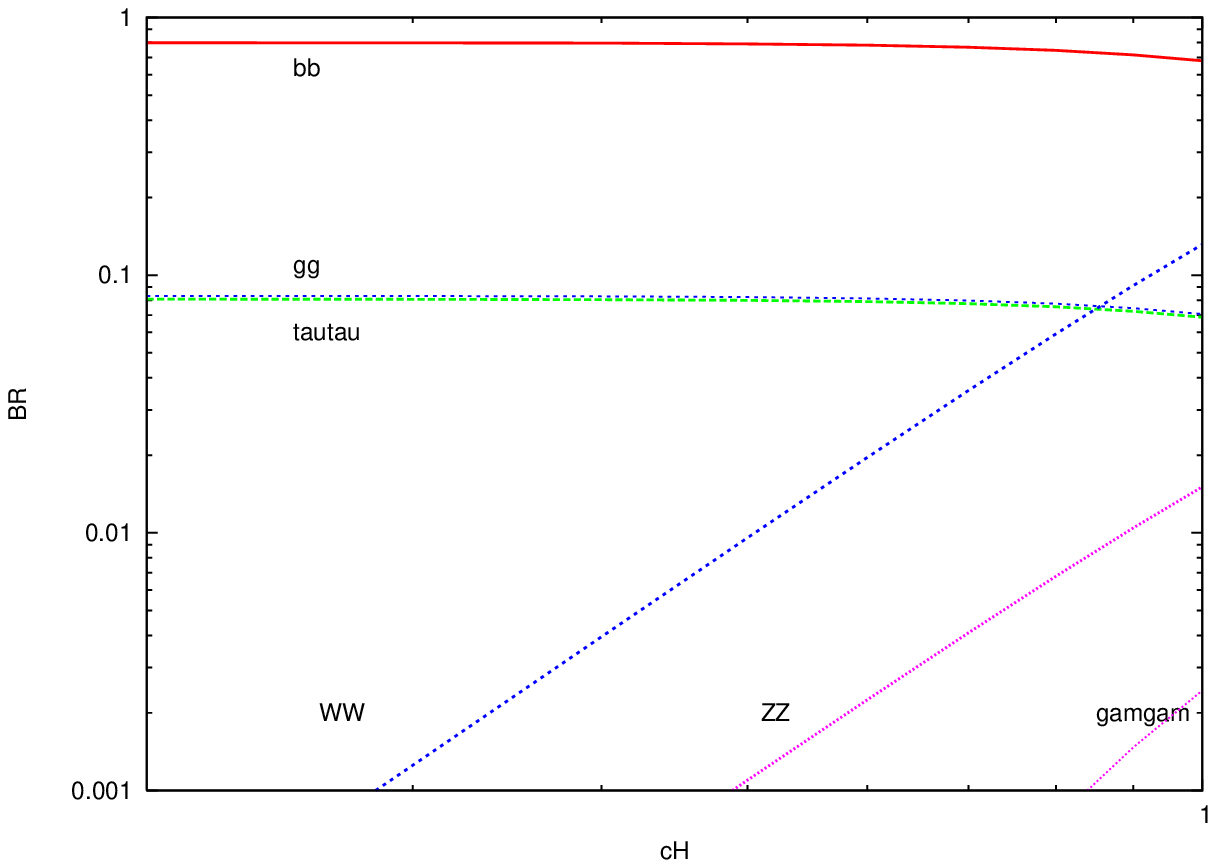}}
	\quad 
	\subfigure[]{\includegraphics[height=5cm,width=.48\textwidth]{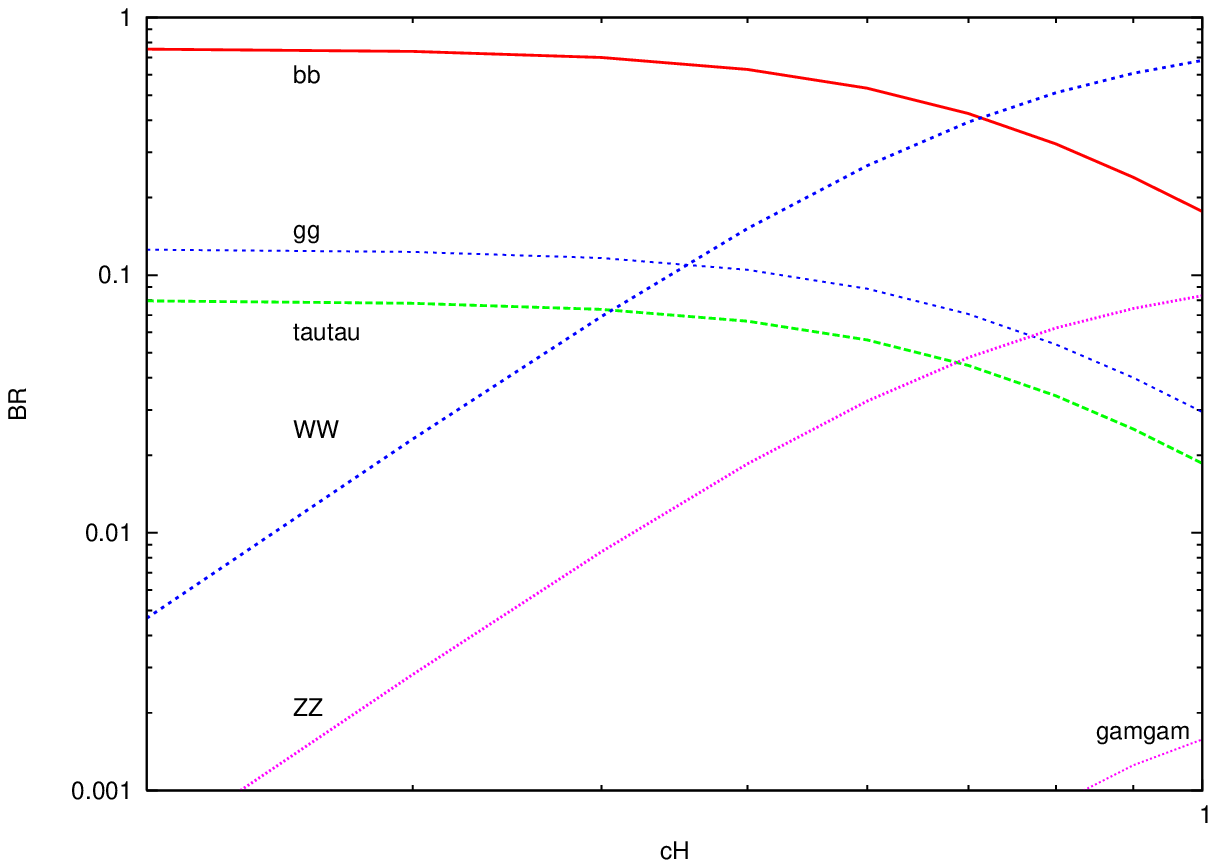}}
 }
\caption{\label{fig:BR}
Branching ratios of $H_1^0$ to the Standard Model particles
for $m_H=120$ GeV (a) and $m_H=150$ GeV (b).
}
\end{figure}

When calculating the branching ratios, it is necessary to consider also
the loop induced decays of the Higgs bosons to gluons and photons.
As mentioned, the tree-level gauge boson couplings to $H_1^0$ are
suppressed by a factor $c_H$.
The $\gamma \gamma H_1^0$ coupling is more
complicated. In the SM, the W-loop gives the dominant contribution 
to the $\gamma \gamma H$ coupling. 
In the triplet model however, 
the fermion coupling is enhanced and the W-coupling suppressed for 
small values of $c_H$, so that the top loop becomes more
important
and
the $\gamma \gamma H_1^0$ coupling is enhanced for small $c_H$. 
However, since the total width increases when $c_H$ decreases,  
the branching ratio to photons remains smaller than in the SM.
The gluon coupling to $H_1^0$ is enhanced by $1/c_H$ due to the
fermion loop.
These effects are seen in Fig.~\ref{fig:BR}, where
the branching ratios of $H_1^0$ are presented for 
$m_{H_1^0}=120$ GeV and 150 GeV.

\section{Central Exclusive Diffractive Production of the Triplet Higgs
Boson \label{sec:cep}}

The central exclusive production (CEP) of a Higgs boson is defined as
$pp\to p \oplus H \oplus p$, where the $\oplus$ denote the 
presence of large rapidity gaps
 between the outgoing protons and the decay products
of the central system. 
It has been suggested in recent years that CEP offers
 a unique complimentary measurement to the
 conventional Higgs search channels, 
 see for example, \cite{KMRpr,Albrow:2008pn}, \cite{DKMOR}~-~\cite{Cox:2007sw}.
Firstly, if the outgoing protons scatter
through small
angles then, to a very good approximation, the primary active di-gluon
system obeys a $J_z=0$, $CP$-even selection rule~\cite{KMRmm}. 
Here $J_z$ is the projection of the total
angular momentum along the proton beam axis.
The observation of the Higgs boson in the CEP
channel therefore determines the Higgs quantum numbers
 to be $J^{PC}=0^{++}$. Secondly, because the process is exclusive, 
 all of the energy/momentum lost by the protons during the
 interaction goes into the production of the central system.
 Measuring the outgoing proton allows the central mass to be 
 measured to just a few GeV, regardless of the decay products of the central system.
 A mass measurement of this type will require 
new forward proton detectors to be installed at ATLAS and/or CMS, 
 which we discuss further in section \ref{sec:fp420}.

For a Standard Model Higgs boson, central exclusive 
diffraction could allow the main
decay channels ($b \bar b$,  $WW$  and $\tau\tau$) to be
observed in the same production channel, which provides the opportunity 
to study the Higgs coupling to $b$-quarks. This may be
very difficult
to access in other search channels at the 
LHC, 
despite the fact that
$H \to b \bar b$ is by far the dominant decay mode for a light SM 
Higgs boson. Furthermore, CEP can provide valuable information on the
 Higgs sector of MSSM, NMSSM and other popular
BSM scenarios  \cite{KKMRext,Ellis:2005fp,hkrstw,Cox:2007sw,fghpp,hkrtw}.
 In this paper, we propose that CEP is also beneficial 
 if higher representations of the Higgs sector are realized, in particular, in searches for the 
Higgs triplets discussed in section \ref{sec:htriplet}.

\begin{figure}
\begin{center}
\includegraphics[height=5cm]{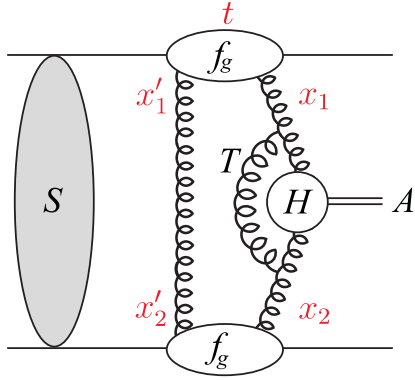}
\caption{A symbolic diagram for the central exclusive production of a Higgs boson $H$.}
\label{fig:parts}
\end{center}
\end{figure}

The theoretical formalism \cite{KMR}~-~\cite{KMRearly}
 for central exclusive production contains distinct parts, as illustrated
in Fig.~\ref{fig:parts}. The cross section can be written
in the form \cite{KMR,KMRpr}
\begin{equation}
\sigma (pp \to p+H+p) ~\sim~\frac{\langle S^2 \rangle}{B^2} \left| \, N\int \frac{dQ_t^2}{Q^4_t}f_g(x_1,x'_1,Q^2_t,\mu^2)f_g(x_2,x'_2,Q^2_t,\mu^2) \right|^2
\label{eq:d1}
\end{equation}
where $B/2$ is the $t$-slope of the proton-Pomeron vertex,
 $\langle S^2 \rangle$ is the soft-survival probability
and the normalization, $N$, is given in terms of the $H\rightarrow gg$ decay width.
The amplitude-squared factor, $\left|...\right|^2$, can be calculated 
in perturbative QCD because the dominant contribution 
 to the integral comes from the 
 region $\Lambda^2_{QCD} \ll Q^2_t \ll m^2_H$ 
 for the 
 large Higgs mass values of interest. 
 The probability amplitudes, 
 $f_g$, to find the appropriate pairs of $t$-channel gluons 
 $(x_1,x'_1)$ and $(x_2,x'_2)$ are given
  by skewed unintegrated gluon densities at a hard scale $\mu \sim m_H/2$.
It is important to emphasize, that  
these  
 generalized gluon distributions
 are usually taken at $p_t=0$, and then the ``total'' exclusive 
 cross section is calculated by integrating over the transverse 
 momentum, $p_T$, of the recoil protons. Assuming an 
 exponential behaviour results in 
\be
\int dp^2_T~e^{-Bp_T^2}~=~1/B~=~\langle p_T^2\rangle.
\ee
Thus, the additional factor in Eq.~(\ref{eq:d1})
is not just the gap survival
but rather the factor  $\langle S^2 \rangle/B^2$
\cite{KMRpr,KMRS}, which
has the form  $S^2\langle p^2_t \rangle^2$ and is much less dependent on
 the parameters of the soft model \cite{KMRS,KMRnns,KMRtalks}

The production cross section for Higgs bosons produced by gluon-gluon fusion and decaying to $b\bar{b}$ is proportional to
\begin{equation}
\frac{\Gamma^{\rm eff}}{m_H^3} \equiv \frac{\Gamma(H\rightarrow gg)}{m_H^3} BR(H\rightarrow b\bar{b})
\end{equation}
where $\Gamma(H\rightarrow gg)$ is the decay width to gluons and BR($H\rightarrow b\bar{b}$) is the branching ratio to $b\bar{b}$ quarks.
Table \ref{tab:cepparams} shows the value of these parameters for the SM Higgs and the lightest Higgs, $H_1^0$, in the triplet model. The central exclusive $H\rightarrow b\bar{b}$ cross section can therefore be enhanced by a large factor with respect to the Standard Model - we discuss this further in Section~\ref{sec:simul}
     
\begin{table}[t]
\centering
\begin{tabular}{|c|c|c|c|c|}
\hline
 & \multicolumn{2}{c|}{$m_H=120$~GeV} & \multicolumn{2}{c|}{$m_H=150$~GeV} \\
& $\Gamma(H\rightarrow gg)$ & BR($H\rightarrow b\bar{b}$) & $\Gamma(H\rightarrow gg)$ & BR($H\rightarrow b\bar{b}$) \\
\hline
$c_H=0.2$ & 6.35$\times$10$^{-3}$ & 0.80 & 1.22$\times$10$^{-2}$ & 0.75 \\
$c_H=0.5$ & 1.01$\times$10$^{-3}$ & 0.79 & 1.95$\times$10$^{-3}$ & 0.63 \\
$c_H=0.8$ & 3.97$\times$10$^{-4}$ & 0.74 & 7.63$\times$10$^{-3}$ & 0.32 \\
SM & 2.49$\times$10$^{-4}$ & 0.68 & 4.79$\times$10$^{-4}$& 0.18\\
\hline
\end{tabular}
\caption{The $H_1^0$ partial decay width to gluons  (expressed in GeV) and branching ratio to $b\bar{b}$ for specific values of Higgs mass and $c_H$. The Standard Model (SM) prediction is shown for comparative purposes. \label{tab:cepparams}}
\end{table}%

The CEP formalism has been extensively checked using the diffractive 
production of $J/\psi$ and the 
leading neutron spectra \cite{JHEP} at HERA and the 
CDF data on central exclusive production processes  ~\cite{CDFdj}~-~\cite{CDFchi}. Further 
tests of the formalism using the early
LHC data have also been suggested \cite{KMRearly}. The main 
uncertainties are associated with:
\begin{itemize}
\item The probability $\langle S^2 \rangle$ that 
additional secondaries will not populate the gaps. 
\item The probability to find the appropriate gluons 
that are given by generalized,
 unintegrated distributions $f_g(x,x',Q_t^2)$.
\item Higher order QCD corrections to the hard subprocess,
 in particular, the  Sudakov suppression. 
\item The so-called `enhanced absorptive corrections' \cite{KKMR,bbkm,KMRnns}
and other effects that may violate the soft-hard factorization.
\end{itemize}

Let us focus 
first
 on the gap survival factor $\langle S^2 \rangle$. 
Since soft physics is involved, we
need a reliable model of soft interactions to quantify the role
of the absorption effects.
In \cite{KMRsoft,KMRnewsoft,KMRnns} soft interaction models were developed and tuned to 
describe all
the available high energy soft $pp$ interaction data. These models
account for
 (i) elastic rescattering (with the two protons in intermediate
states), (ii) the probability of low-mass proton excitations,
and (iii) the screening corrections due to high-mass proton
dissociation (enhanced absorption). The first two effects result
in the so-called eikonal contribution,  $\langle S^2_{\rm eik} \rangle$.
In the most recent version of the soft rescattering model \cite{KMRnns,KMRtalks}, the 
KMR group
obtained $\langle S^2_{\rm eik} \rangle_{\rm eff}=0.025$ when 
adjusting  $\langle S^2_{\rm eik} \rangle$
to its value corresponding to an exponential slope $B=4$ GeV$^{-2}$. 


In the presence of enhanced screening, however, there is no longer exact factorisation
  between the hard and soft parts of the process, see for example \cite{bbkm,JHEP,KMRnns}. 
The latest calculations \cite{KMRnns} indicate that, 
in the case of the SM Higgs production
at the LHC,  the effective survival
factor, due to both eikonal 
and enhanced rescattering, is  
$\langle S^2 \rangle_{\rm eff} \simeq 0.015 \pm 0.01$.
It should be noted that the exclusive dijet, $\gamma \gamma$ and $\chi_c$ production data from CDF
 and the leading neutron data at HERA indicate that
$\langle S^2_{\rm enh} \rangle$ is somewhat larger
  than this estimate, such that $\langle S^2 \rangle_{\rm eff}$ 
 is nearer the upper limit of the quoted interval. 
In any case, it will be possible to measure 
$\langle S^2_{\rm enh} \rangle$ using early LHC data \cite{KMRearly,Orava}.

As the generalized, unintegrated gluon distribution $f_g$ has not been
measured explicitly, it is 
obtained in the KMR approach \cite{KMR,KMRpr} from the conventional 
gluon distribution,  $g(x,Q_t^2)$,
 known from the global parton analyses.
The main uncertainty comes from the lack of knowledge 
of the integrated gluon distribution at low $x$ and small scales.
It was found in \cite{KMRearly}
that  a variety of recent global analyses give a spread of
\be
xg~=~(3-3.8)~~{\rm for}~~x=10^{-2}~~~~{\rm and}~~~~xg~=~(3.4-4.5)~~{\rm for}~~x=10^{-3}
\ee
for $Q_t^2=4$~GeV$^2$. 
These are big uncertainties bearing in mind that the CEP cross section 
depends on $(xg)^4$. A similar estimate of the uncertainty from the input gluon 
distribution functions was presented in \cite{Cox:2007sw}. The uncertainty related to the
Sudakov factor is addressed in \cite{KMRearly}, along with the
 measurements that could reduce the uncertainty using early LHC data.

The overall
uncertainty factor in the calculation of the
CEP of Higgs bosons at the LHC
was estimated to be approximately
2.5 in \cite{KKMRext,hkrstw}.
Again, we note that the first LHC runs will allow the accuracy to be drastically improved.

\section{Simulation of Higgs production in the triplet model\label{sec:simul}}

\subsection{Forward proton tagging\label{sec:fp420}}

The forward proton detectors will need to be installed in the high dispersion
regions 220 m 
and 420 m either side of the interaction point at ATLAS and/or 
CMS\footnote{We refer the reader to the FP420 R\&D report for 
more details on the detectors \cite{Albrow:2008pn}.}. 
We restrict our discussion and analysis to the ATLAS interaction 
point (IP), and note that a similar result would be obtained at CMS.
There are three important aspects of forward proton tagging at the 
LHC that need to be considered for the purposes of this analysis; 
the acceptance and resolution of the proposed forward proton 
detectors and the ability of the detectors to measure the 
time-of-flight of each proton from the interaction point.

The acceptance of the forward proton system depends on the distance 
that each active detector is from the LHC beam. 
We choose the proton detectors located at 220 m (420 m) from the 
IP to be 2 mm (5 mm) from the beam and use the FPtrack program 
\cite{fptrack} to track the path of the protons through the LHC 
lattice in order to fully determine the acceptance. 
The acceptance of the forward proton detector system is dependent 
on the mass of the centrally produced object, $M$, which is given by
\begin{equation}
M^2 = \xi_1 \, \xi_2 \, s
\end{equation}
where $\xi_1$ and $\xi_2$ are the fractional longitudinal momentum
losses of the outgoing protons. 
For central masses less than 200~GeV, the protons are either both 
detected at 420m (symmetric tagging) or one is detected at 220 m and 
one at 420 m (asymmetric tagging). For example, the acceptance for a 
central mass of 120~GeV is approximately 28\% for symmetric events 
and 20\% for asymmetric events\footnote{Dead material in the detectors 
at 220 m can 
reduce the acceptance for symmetric tagging. We do 
not consider that effect here.} \cite{Albrow:2008pn}.
 This acceptance 
changes to 20\% and 40\% respectively for a 150~GeV central mass.
Furthermore, symmetric and asymmetric events also have different mass 
resolution for a given central mass; the resolution of a 150~GeV Higgs 
boson is approximately 1.5~GeV and 4~GeV for symmetric and asymmetric 
proton tagging respectively.

Finally, the forward proton detectors will be capable of measuring the 
time-of-flight of each proton from the interaction point to an accuracy 
of 10~ps. The difference in the time-of-flight measurement of the 
protons gives a measurement of the interaction vertex to 2.1 mm, 
assuming that the reference timing system has negligible jitter. 
This vertex reconstruction proves to be very useful in background 
rejection, as discussed in Section \ref{sec:expcuts}.

\subsection{Signal and background event generation\label{sec:evgen}}

The central exclusive signal and background events are simulated with 
full parton showering and hadronization effects using the ExHuME v1.3.4 
event generator \cite{Monk:2005ji}. 
ExHuME contains a direct  implementation of the KMR calculation \cite{KMR,KMRpr}
of central exclusive diffraction given in Sec. \ref{sec:cep}. The CTEQ6M 
\cite{Pumplin:2002vw} parton distribution functions are used to calculate 
the generalized gluon distributions, $f_g$. 
The generator level cross sections for central exclusive 
$H\rightarrow b\bar{b}$ production in the triplet model are presented 
in Table \ref{tab:genxs} for $m_H=120,150$~GeV and $c_H=0.2,0.5,0.8$. 
In Section \ref{sec:res} we will present results for each of these 
parameter choices.

\begin{table}[t]
\centering
\begin{tabular}{|c|c|c|}
\hline
$\sigma_{H\rightarrow b\bar{b}}$~(fb) & $m_H=120$~GeV & $m_H=150$~GeV \\
\hline
$c_H=0.2$ & 113.5 & 55.2 \\
$c_H=0.5$ & 18.0 & 7.4 \\
$c_H=0.8$ & 6.6 & 1.5 \\
\hline
\end{tabular}
\caption{Generator level cross sections, $\sigma_{H\rightarrow
b\bar{b}}$,
for central exclusive Higgs boson production for $m_H=120,150$~GeV
and $c_H=0.2,0.5,0.8$. \label{tab:genxs}}
\end{table}%

The backgrounds to $H\rightarrow b\bar{b}$ can be broken down into three
broad categories; central exclusive, double pomeron exchange and
overlap.
We use the ExHuME event generator for the central exclusive backgrounds.
 ExHuME contains the leading order calculation for central exclusive
$b\bar{b}$ production. Recent results however \cite{Shuvaev:2008yn},
show that the one-loop corrections to this process reduce the
cross section by a factor of approximately two. We normalize the
$b\bar{b}$ events generated by ExHuME accordingly.
Central exclusive $gg$ production has a much larger cross section
than $b\bar{b}$ and can act as a background when the gluon jets are
mis-identified by the b-tagging algorithms.
At ATLAS, the mis-identification rate for each gluon jet is 1.3\%
for a b-tagging efficiency of 60\% \cite{btag}.
We do not generate exclusive $c\bar{c}$ events for two reasons.
Firstly, due to the $J_z=0$ spin-selection rule \cite{KMRmm}, exclusive
$c\bar{c}$ production is suppressed with respect to $b\bar{b}$
by a factor of $m_c^2/m_b^2$.
Furthermore, the mis-tag rate for c-jets to be identified as b-jets
is $\sim$10\%.
Thus, the background contribution from exclusive $c\bar{c}$ events
is considered to be negligible in comparison to the exclusive
$b\bar{b}$ events.
Higher order events, such as $b\bar{b}g$ have been studied in
\cite{Khoze:2006um} with the conclusion that these types of
events should be negligible after all experimental cuts.
The demonstration of this using event generators cannot be
completed until the relevant processes are implemented into
the ExHuME Monte Carlo.

Double pomeron exchange (DPE) is the process
$pp \rightarrow p + X + p$, where the central system, $X$, is
produced by pomeron-pomeron fusion.
The pomeron is assigned a partonic structure and so there are
always `pomeron remnants' accompanying the hard scatter.
DPE has been extensively studied in relation to
$H\rightarrow b\bar{b}$ and it has been concluded that this 
background is negligible after appropriate experimental cuts
\cite{Cox:2007sw,Khoze:2007hx}.
We do
not consider these types of events further.

In addition to these standard backgrounds, we also examine the
effect of the overlap backgrounds.
This source of background is important when there are a large number
of $pp$ interactions in each bunch crossing at the LHC. The largest
overlap background is a three-fold coincidence between two soft
 events ($pp\rightarrow pX$), which produce forward protons within
the acceptance of the forward detectors, and an inelastic event, which
produces the hard scatter $pp\rightarrow X$ and,
thus, can mimic our signal.
 To simulate these events we use HERWIG \cite{Corcella:2002jc} plus
JIMMY \cite{Butterworth:1996zw} to generate $pp\rightarrow b\bar{b}$,
using the tune (A) to Tevatron data \cite{jimmytune}.
The forward protons are then added into the event using the prescription
given in \cite{Cox:2007sw}, which also allows us to calculate the
probability of the coincidence as a function of the LHC luminosity.

The overlap background is initially reduced using the vertex matching
provided by the proton time-of-flight (TOF) information.
As the protons do not come from the same interaction as the jets, the
vertex reconstructed using TOF will not, in general, coincide with the
di-jet vertex. Given a fast-timing resolution of 10 ps, a rejection
factor of 18 (15) can be obtained at low (high) luminosity by requiring
that the di-jet vertex be within $\pm$4.2mm of the `fake' vertex
reconstructed from TOF.
Approximately 95\% of the CEP events will be retained by this
requirement.

We do not consider the backgrounds from two-fold coincidences.
It was demonstrated in \cite{Cox:2007sw} that the largest of these
backgrounds - the coincidence between a soft central diffractive
scattering and a standard QCD $2\rightarrow2$ scattering - was at
least a factor of five smaller than the threefold coincidence.
Furthermore, as discussed in the FP420 R\&D report \cite{Albrow:2008pn},
this background is (i) probably overestimated and (ii) will be
additionally rejected by the charged track cut outlined in the
next section.

When generating the event samples, we require that the central mass be
in the range $80<M<250$ GeV, which improves the event generator
efficiency and is the broad region of interest for this study.
To approximate the detector effects, we smear the energy, momenta
and angles of each central final state particle according to the
ATLAS detector resolution \cite{calorimeters}.
The outgoing forward proton momenta are smeared by the amount
specified in \cite{Bussey:2006vx}.
A mid-point cone algorithm is then applied to the samples and events
retained if the leading jet has transverse energy greater than 45~GeV.
Finally, b-tagging efficiencies are imposed after matching the
two leading jets to the partonic level.

\subsection{Experimental cuts\label{sec:expcuts}}

To enhance the signal, we follow the experimental method used in a 
previous study of $H\rightarrow b\bar{b}$ in the SM and the 
MSSM \cite{Cox:2007sw}, which imposes a number of exclusivity 
cuts\footnote{A somewhat different experimental method was discussed 
in \cite{hkrstw}. However, the final experimental efficiency broadly agrees 
with that used here \cite{Albrow:2008pn}.}.
Firstly, the rapidity of the central system can be estimated from 
the forward proton detectors by
\begin{equation}
y=\frac{1}{2}\rm{ln}\left(\frac{\xi_1}{\xi_2} \right).
\end{equation}
The difference, $\Delta y$, between this rapidity measurement and 
the average pseudo-rapidity of the di-jets should be approximately 
zero for an exclusive event, Exclusive candidates satisfy
\begin{equation}
\Delta y = \left| y - \left( \frac{\eta_1 + \eta_2}{2} \right) 
\right| \leq 0.06,
\end{equation}
where $\eta_1$ and $\eta_2$ are the pseudo-rapidities of the leading
jets.

The di-jet mass fraction, $R_j$, determines the fraction of the 
central mass that is contained within the di-jet system. $R_j$ is an
improved version of the $R_{jj}$ variable \cite{Khoze:2006iw}, 
which was used by CDF in the search for exclusive di-jet events
\cite{CDFdj}.
For an exclusive di-jet event, one expects all of the mass to be
contained in the di-jets, and hence $R_j=1$.
However, parton showering/hadronisation effects can result in
energy outside of the jets. Furthermore, detector resolution will
smear the di-jet mass measurement. An exclusive event is defined to be
\begin{equation}
0.85 \quad \leq \quad R_j = \frac{2E_T^1}{M}\rm{cosh}\left(
\eta_1 - y \right) \quad \leq \quad 1.1,
\end{equation}
where $E_T^1$ is the transverse energy of the leading jet.

The third exclusivity cut requires the di-jets to be back to back in 
azimuth, which reflects the suppression of initial state radiation 
for an exclusive di-jet event. The back to back requirement is
\begin{equation}
|\pi -\Delta\phi| \leq 0.15,
\end{equation}
where $\Delta \phi$ is the azimuthal angle between the jets. It is also 
possible to examine each event for underlying event activity, caused 
by multiple parton-parton interactions. 
The exclusive events do not have these additional scatters because 
the proton remains intact. It is possible to reject inclusive events, 
and, hence, the overlap background, by requiring few charged tracks,
$N_C$, associated with the di-jet vertex but outside of the jets.
This definition is, of course, dependent on the jet algorithm used to
define the jets. An algorithm independent approach is to examine the
charged track activity perpendicular to the leading jet, $N_C^{\perp}$.
In this approach, charged tracks are assigned to the underlying event
if they satisfy
\begin{equation}
\frac{\pi}{3} \leq | \phi_k - \phi_1 | \leq \frac{2\pi}{3} \quad
\textrm{or} \quad
\frac{4\pi}{3} \leq | \phi_k - \phi_1 | \leq \frac{5\pi}{3},
\end{equation}
where $\phi_k$ is a the azimuthal angle of a charged track and $\phi_1$ 
is the azimuthal angle of the leading jet. 
In this analysis, we identify exclusive events by
\begin{equation}
N_C \leq 3 \quad \rm{and} \quad N_C^{\perp} \leq 1.
\end{equation}

For completeness, in Table  \ref{tab:finalxs} we present the final cross sections 
for the signal 
($m_H=$120~GeV, $c_H=0.5$) and background events.
It should be noted that the signal is concentrated at $M=120$~GeV,
whereas the backgrounds form a continuum across the mass range
80$<$M$<$250~GeV.
For details on the efficiencies of the individual cuts, and
motivation for the cut choices, we refer the reader to
\cite{Cox:2007sw}.
The overlap backgrounds are luminosity dependent and are presented
for constant luminosities of 10$^{33}$~cm$^{-2}$~s$^{-1}$ (low) and
10$^{34}$~cm$^{-2}$~s$^{-1}$ (high).

\begin{table}[t]
\centering
\begin{tabular}{|c|c|c|c|}
\hline
Generator & Process & $\sigma_{420-420}$ (fb) & $\sigma_{420-220}$ (fb)
\\
\hline
ExHuME & $H\rightarrow b\bar{b}$ & 0.53 & 0.28 \\
\hline
& $b\bar{b}$ & 0.53 & 0.27 \\
 & $gg$ & 1.08 & 0.91 \\
 \hline
Overlap (L) & $b\bar{b}$  & 0.07 & 0.09 \\
Overlap (H) & $b\bar{b}$ & 11.0 & 13.7 \\
\hline
Total bgrd (L)  & & 1.68 & 1.27 \\
Total bgrd (H) & & 12.6 & 14.9 \\
\hline
\end{tabular}
\caption{The final cross sections for the $H\rightarrow b\bar{b}$
($m_H=120$~GeV, $c_H=0.5$) and relevant backgrounds after the all 
cuts discussed in the text. 
The overlap backgrounds are defined at both low (L) and high (H)
luminosity. All the backgrounds form a continuum over the range
80$<$M$<$250~GeV.  \label{tab:finalxs}}
\end{table}

\subsection{Trigger strategies\label{sec:trigger}}

A major experimental challenge  for central exclusive jet analyses is
developing a trigger strategy to retain enough events. 
At ATLAS, jets with $E_T\sim50$~GeV are heavily prescaled in the level
one (L1) trigger, due to the high rate and the lack of additional
rejective power in the high level trigger (HLT).
The total L1 rate allowed at ATLAS is 75-100 kHz, which must be
reduced to $\sim$100~Hz after the HLT.
In this section, we discuss three possible trigger strategies.
The first  possibility is to exploit the muon rich nature of 
$b\bar{b}$ events. 
The lowest muon threshold (MU6) at ATLAS is designed to retain 80\% of
muons with transverse momentum greater than 6~GeV.
The single muon trigger efficiency for $b\bar{b}$ events is
11\% \cite{Cox:2007sw}. In order to keep the L1 rate down it
will be necessary to require that the event contains jet with
$E_T>40$~GeV in conjunction with the muon.

The second trigger strategy is to require a 40~GeV jet in conjunction 
with a proton tagged in a detector at 220 m from the interaction 
point\footnote{The information from the detectors at 420m will not
reach the central trigger processor within the
latency of 2.5$\mu$s and so cannot be used in the L1 decision.}. 
This trigger has been extensively studied in previous work
\cite{Grothe:2006dj}, and it is expected that the unprescaled L1 rate
will be less than 1 kHz up to a luminosity of
L$=$2$\times$10$^{33}$~cm$^{-2}$~s$^{-1}$.
This rate however, will scale with L$^2$ and the trigger may have
to be prescaled to give a fixed rate at the highest luminosity.
We will investigate two fixed rate triggers; R5 is a L1 rate of
5 kHz and R10 is a L1 rate of 10 kHz.
The drawback of this trigger strategy is that the symmetric events
will not be retained, which could potentially have a large impact
for a light Higgs boson measurement.

It will be possible to dramatically reduce any high L1 rate using 
additional information in the HLT.
Firstly, requiring that there be a proton detected at 420 m and using
time-of-flight information to match the vertices will reject the
events by a factor of approximately 600 (60) at low (high) luminosity.
A loose b-tagging requirement could also be added.
Finally, the $R_j$ and $\Delta y$ cuts reject the overlap  events by
a factor of approximately 100 and could be used to further reduce
the rate.

The final trigger strategy is to allow a larger L1 rate for the 
40~GeV jets, which is then reduced in the HLT by requiring two in-time
protons. This trigger has also been studied in previous work
\cite{Cox:2007sw}. 
The advantage of this approach is that it would retain both symmetric
and asymmetric events.
The L1 rate for 40~GeV jets is expected to be approximately 25~kHz at
low luminosity, rising to 250kHz at high 
luminosity. We define a trigger for 40~GeV jets, JR25, which has a fixed
rate of 25~kHz\footnote{It was shown in CMS-based study \cite{CMS-Totem} that this rate 
can be reduced by a factor of two by requiring that
the majority of the transverse energy in the detector be contained
within the dijets - i.e. that $(E_T^1 + E_T^2)/H_T > 0.9$, where
$H_T$ is the scalar sum of transverse energy deposited in the  
detector \cite{Grothe:2006dj}.}. This trigger would be unprescaled at low
luminosity and prescaled by a factor of ten at high luminosity.

\subsection{Significance of observation and expected mass
distributions \label{sec:res}}

In this section, we estimate the significance of observing a neutral
Higgs
boson in the triplet model for the parameter choices presented in
section \ref{sec:evgen} and for the trigger strategies outlined in
section \ref{sec:trigger}.
As the overlap background is luminosity dependent (Table
\ref{tab:finalxs}),
we must specify how the data was collected.
For example, we examine 
the significance for an integrated luminosity of 60~fb$^{-1}$, which 
corresponds to between three and four years of data acquisition given  
a  peak luminosity of 2$\times$10$^{33}$~cm$^{-2}$~s$^{-1}$. 
We also present results for 300~fb$^{-1}$ of data, which corresponds 
to between three and four years of data acquisition given a peak 
luminosity of 10$^{34}$~cm$^{-2}$~s$^{-1}$. 
It should be noted, however, that the data acquisition at the LHC 
will not be collected at one specific luminosity. 
Firstly, the peak luminosity will increase during the lifetime of 
the LHC with an improved understanding of the machine. 
Secondly, during each store, the luminosity will decrease exponentially 
with a lifetime of approximately 14~hours. 
We crudely approximate these effects by defining that half the 
data is collected at the peak luminosity and half the data at 
75\% of the maximum.

The signal and background events that pass the selection criteria are
normalized to an expected number, $N$, for each trigger/luminosity
scenario, i.e.
$$N=L \sigma \epsilon,$$
 where $L$ is the total integrated luminosity, $\sigma$ is the final 
cross section for each process as shown in Table \ref{tab:finalxs} 
and $\epsilon$ is the trigger efficiency. In the case of the overlap 
background, $\sigma$ is dependent on the assumed peak luminosity as 
described above.
The MC distributions are then used to predict the expected number
of events in each bin of a mass distribution.
A pseudo-experiment is then carried out by randomly picking a number
of events for each bin in the mass distribution according to a
Poisson distribution.
The significance of each pseudo-experiment is then obtained by fitting
the pseudo-data with a signal-plus-background function and a
background-only function.
The significance is estimated from the difference in $\chi^2$ of the
two fits by
\begin{equation}
S = \sqrt{\chi_b^2 - \chi_{s+b}^2}.
\end{equation}
We assume the background function will be well known from data - in this
analysis we use the weighted MC background events to determine the 
shape and allow the normalization to vary\footnote{We have checked our
results using a quadratic background function and observe little
difference.}. The signal function is a Gaussian and all parameters
are allowed to vary.
We repeat the procedure for 500 pseudo-experiments to determine the
average significance of each luminosity/trigger scenario and ensure
that our results are consistent and the presented distributions typical.

\begin{figure}
\centering
\mbox{
       
\subfigure[]{\includegraphics[width=.5\textwidth]{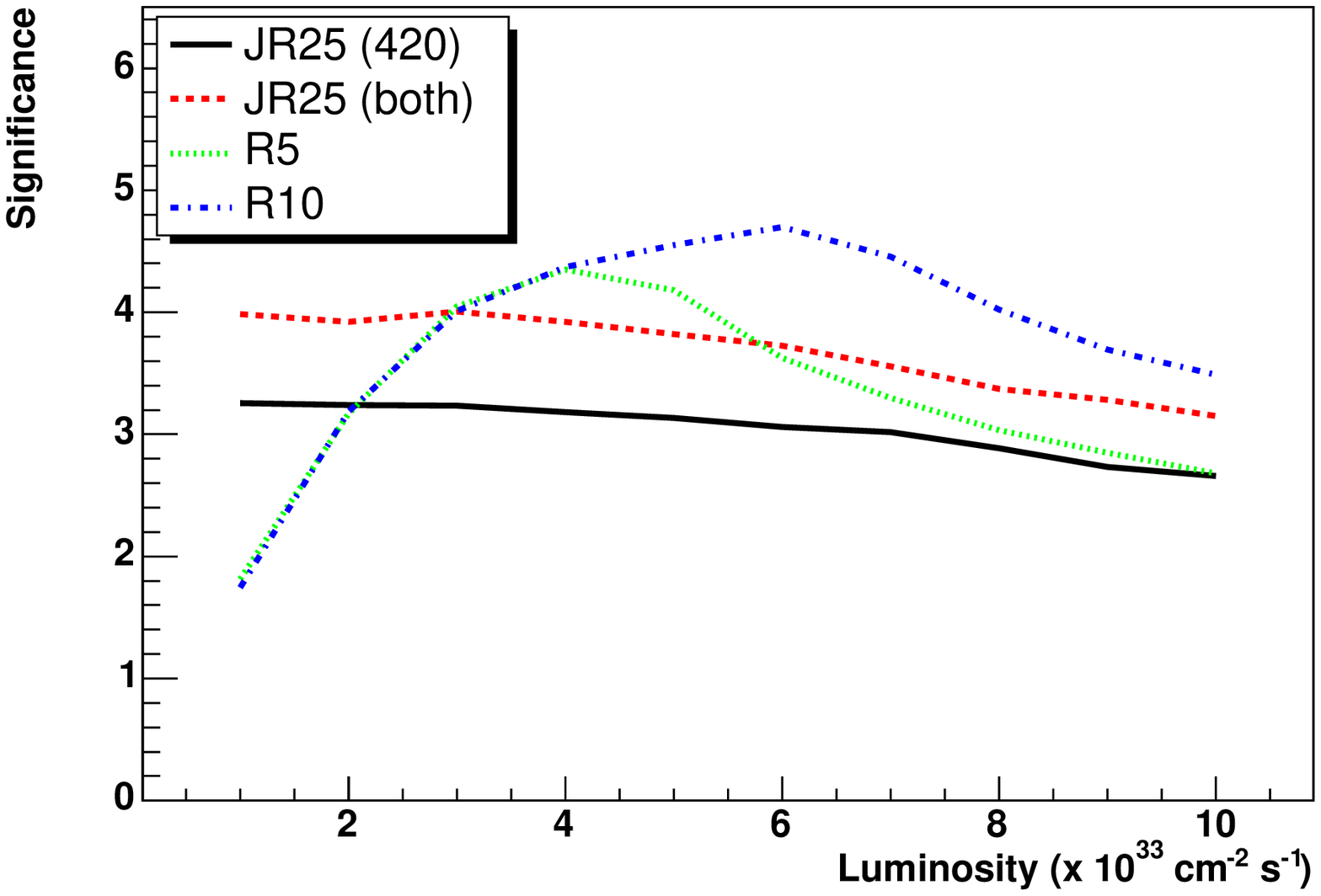}}
\quad
       
\subfigure[]{\includegraphics[width=.5\textwidth]{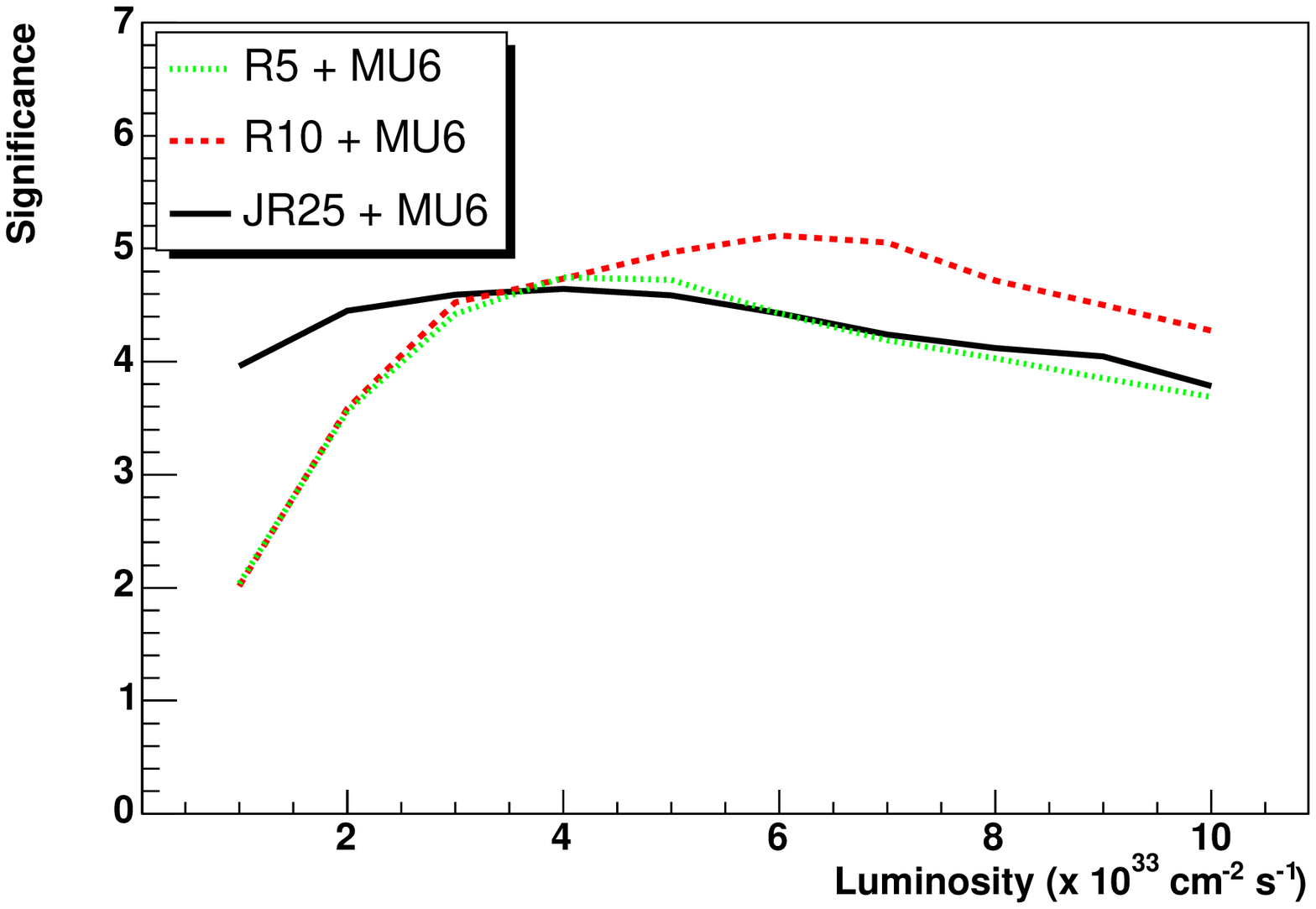}}
        }
\caption{(a) Expected significance ($m_H=120$~GeV, $c_H=0.5$) for the
R5, R10 and JR25 triggers as a function of luminosity given three years
of data acquisition at each luminosity. We assume that 10~fb$^{-1}$ of 
data is collected each year per 10$^{33}$~cm$^{-2}$~s$^{-1}$ of 
luminosity, i.e. 120~fb$^{-1}$ of data is collected at 
4$\times$10$^{33}$~cm$^{-2}$~s$^{-1}$. (b) The significance of the 
R5, R10 and JR25 triggers when combined with the MU6 trigger. 
\label{fig:trigrates}}
\end{figure}

To determine the significance for each parameter choice, the trigger 
strategies outlined in Sec \ref{sec:trigger} are evaluated and the best
method chosen to retain the events.
We choose the $m_H=120$~GeV, $c_H=0.5$ point as our reference.
Figure \ref{fig:trigrates} (a) shows the significance as a function of
luminosity for the R5, R10 and JR25 trigger strategies given three
years of data acquisition at that luminosity.
It is clear that the JR25 trigger is the best choice at low luminosity
as it retains a high fraction of both symmetric and asymmetric events.
At higher luminosities, the R5/R10 triggers become more favourable;
the R5 (R10) trigger do not become prescaled until
4.5$\times$10$^{33}$~cm$^{-2}$~s$^{-1}$
(6.3$\times$10$^{33}$~cm$^{-2}$~s$^{-1}$) and the number of exclusive
events in the final sample is increased even though the symmetric
events are not retained. Figure \ref{fig:trigrates} (b) shows the
significance for R5, R10 and JR25 when combined with the MU6 trigger.

\begin{figure}
\centering
\mbox{
       
\subfigure[]{\includegraphics[width=.5\textwidth]{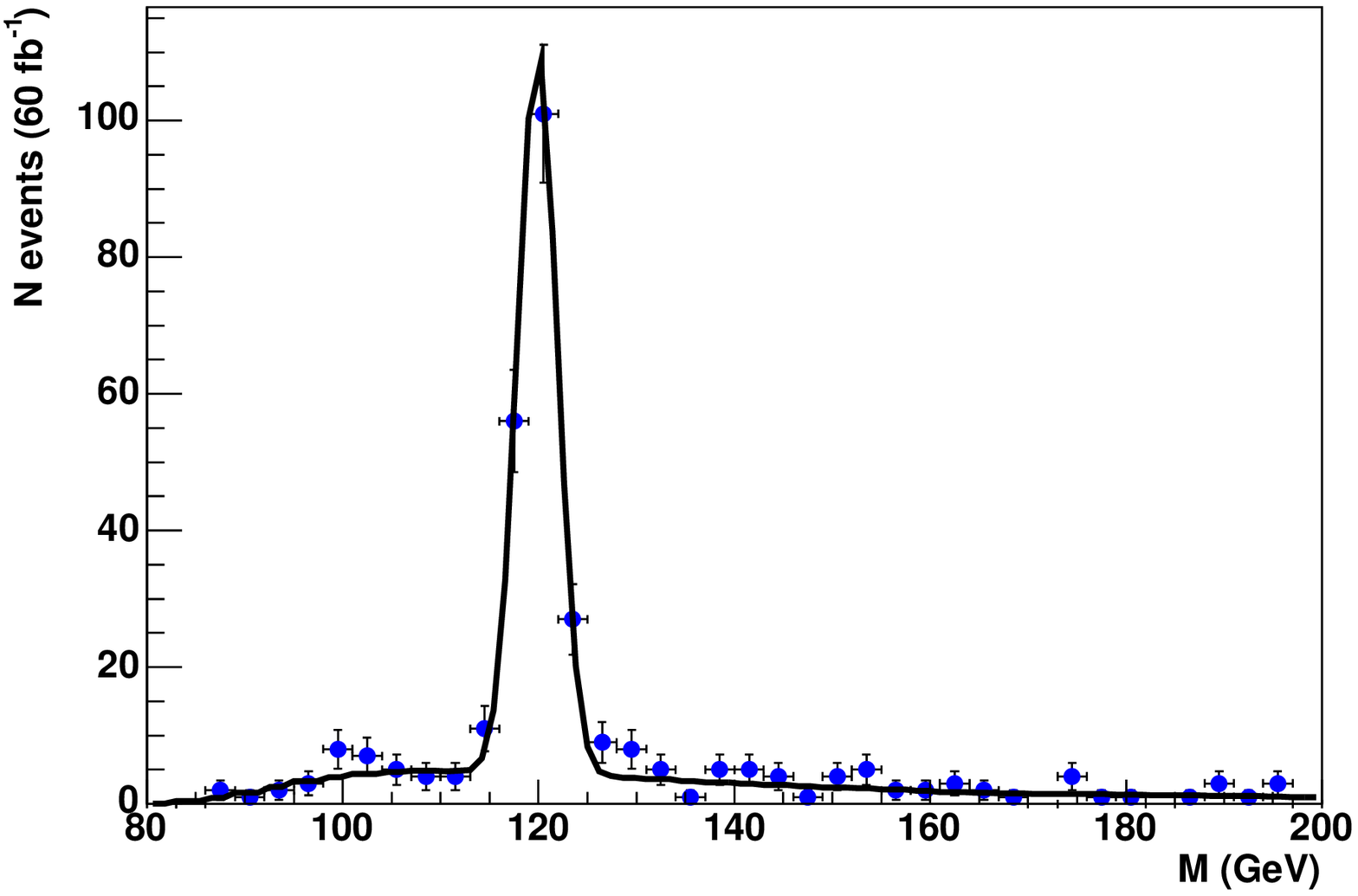}}
\quad
       
\subfigure[]{\includegraphics[width=.5\textwidth]{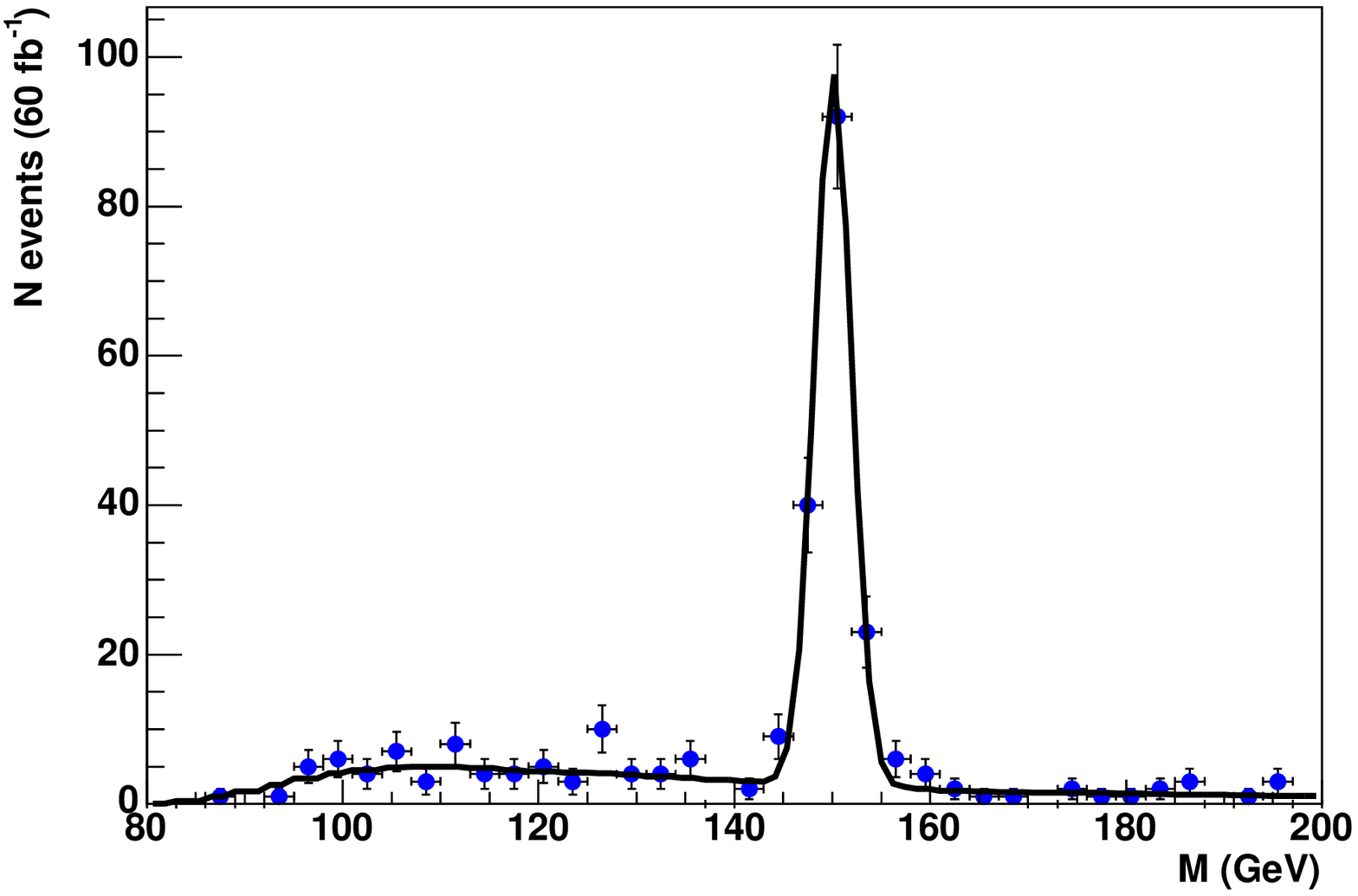}}
        }
\mbox{
       
\subfigure[]{\includegraphics[width=.5\textwidth]{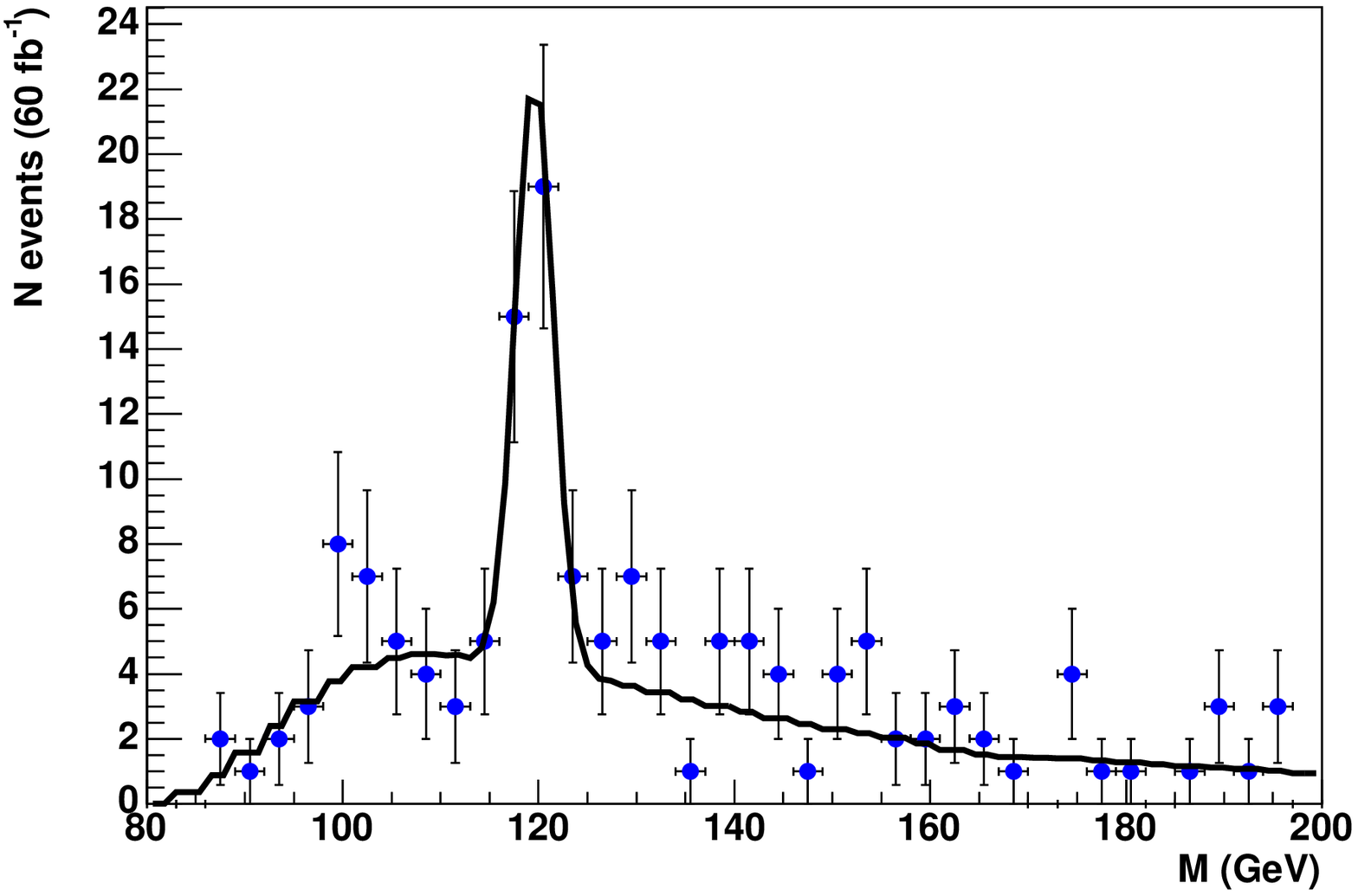}}
\quad
       
\subfigure[]{\includegraphics[width=.5\textwidth]{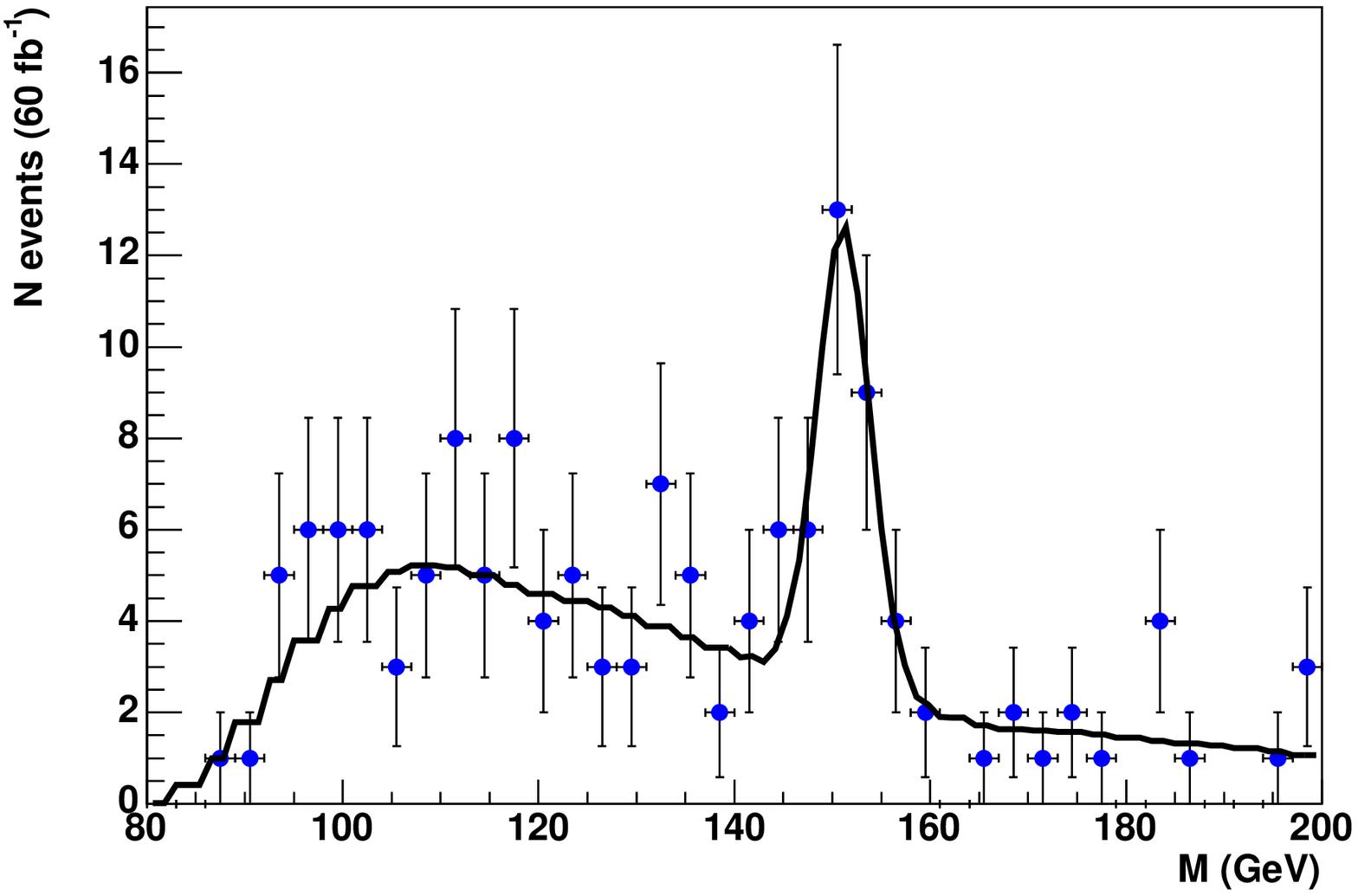}}
        }
\caption{Expected mass distributions given 60~fb$^{-1}$ of data,
collected at 2$\times$10$^{33}$~cm$^{-2}$~s$^{-1}$ using a JR25$+$MU6 
trigger, for the following parameter choices: (a) $m_H=120$~GeV 
and $c_H=0.2$, significance is 12.7$\sigma$ (b) $m_H=150$~GeV 
and $c_H=0.2$, significance is 11.9$\sigma$.  
(c) $m_H=120$~GeV and $c_H=0.5$, significance is 4.5$\sigma$.
(d) $m_H=150$~GeV and $c_H=0.5$, significance is 3.9$\sigma$.
\label{fig:lowlumistandard}}
\end{figure}

\begin{figure}
\centering
\mbox{
       
\subfigure[]{\includegraphics[width=.5\textwidth]{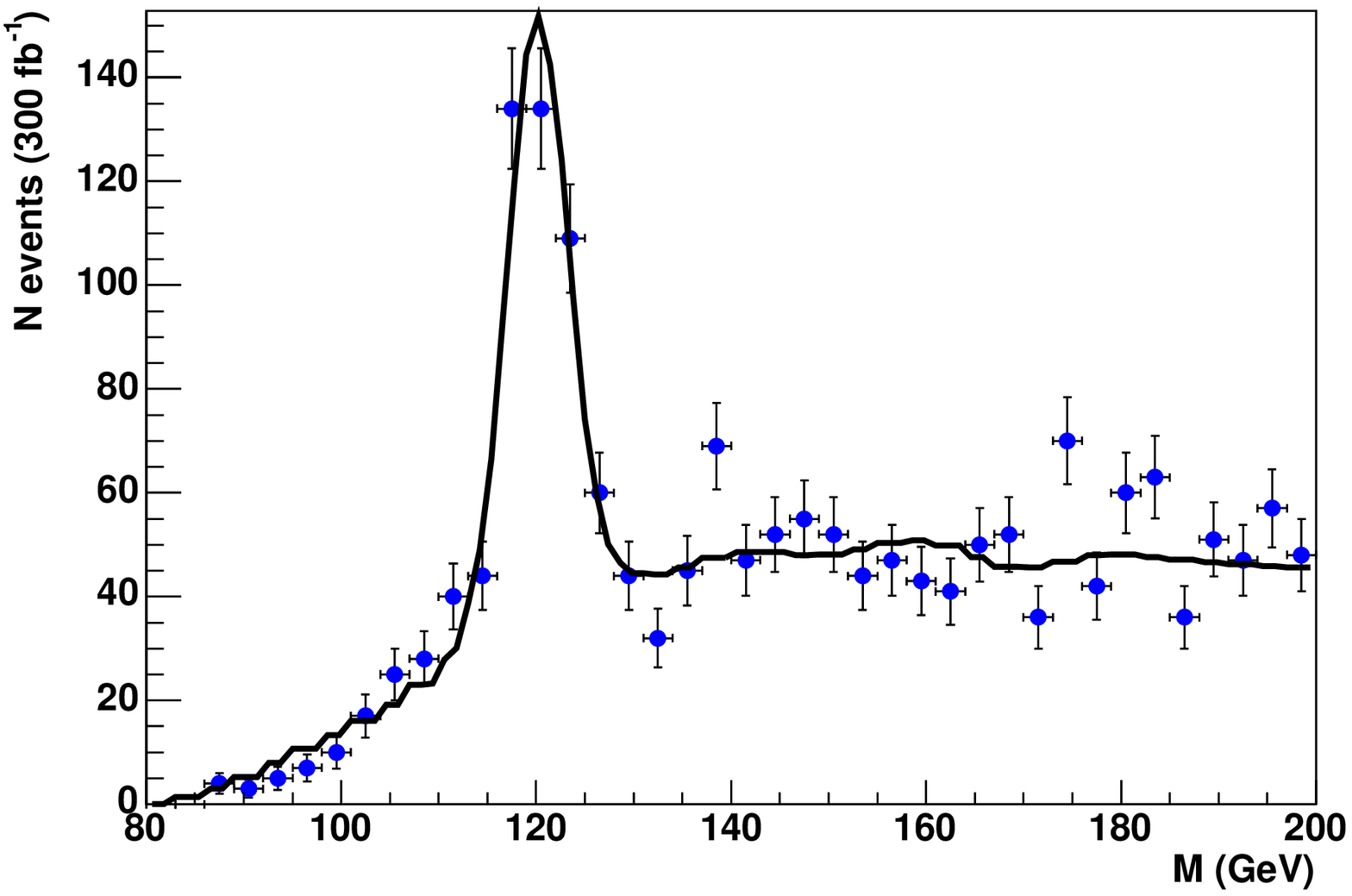}}
\quad
       
\subfigure[]{\includegraphics[width=.5\textwidth]{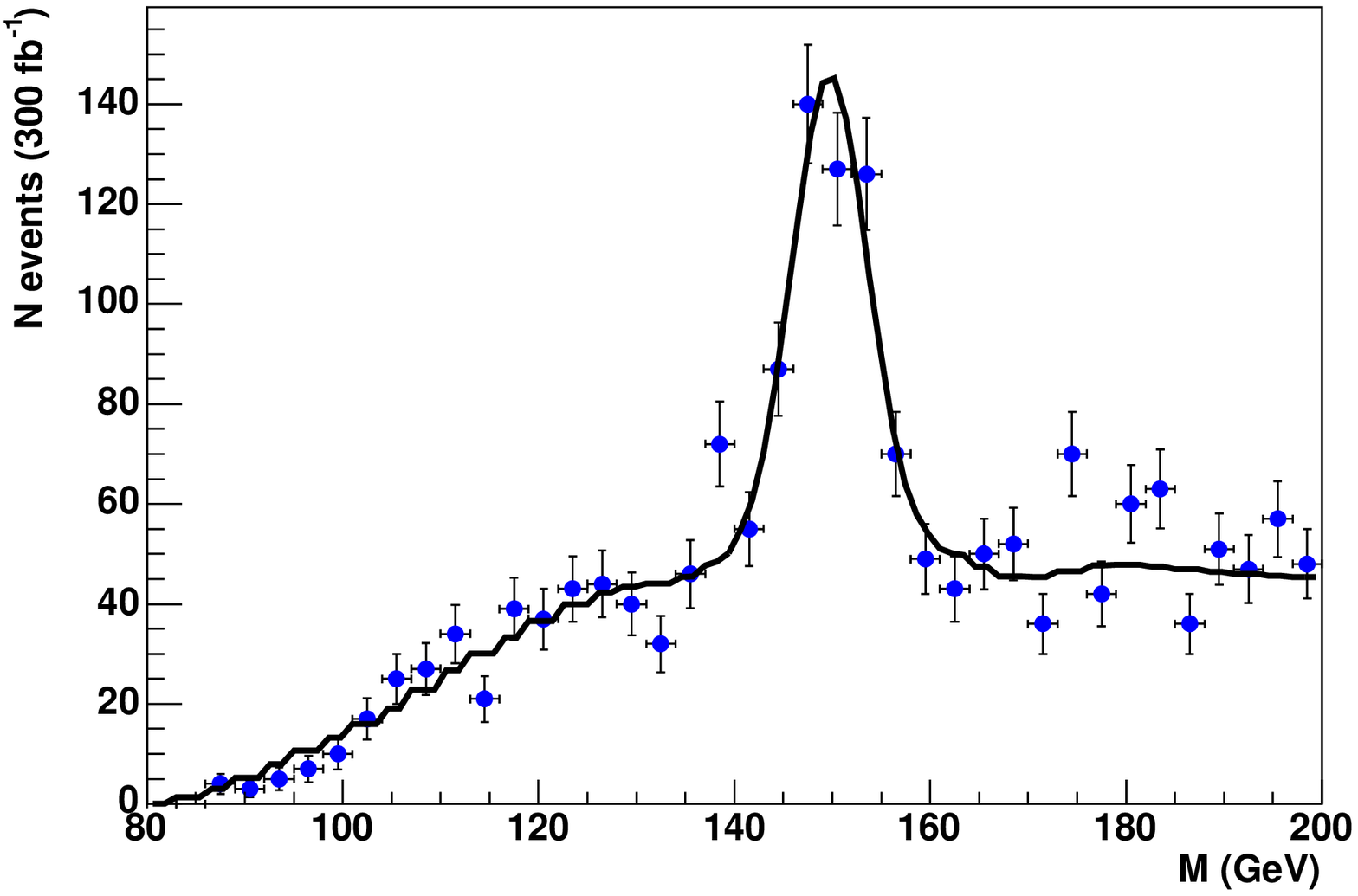}}
        }
\mbox{
       
\subfigure[]{\includegraphics[width=.5\textwidth]{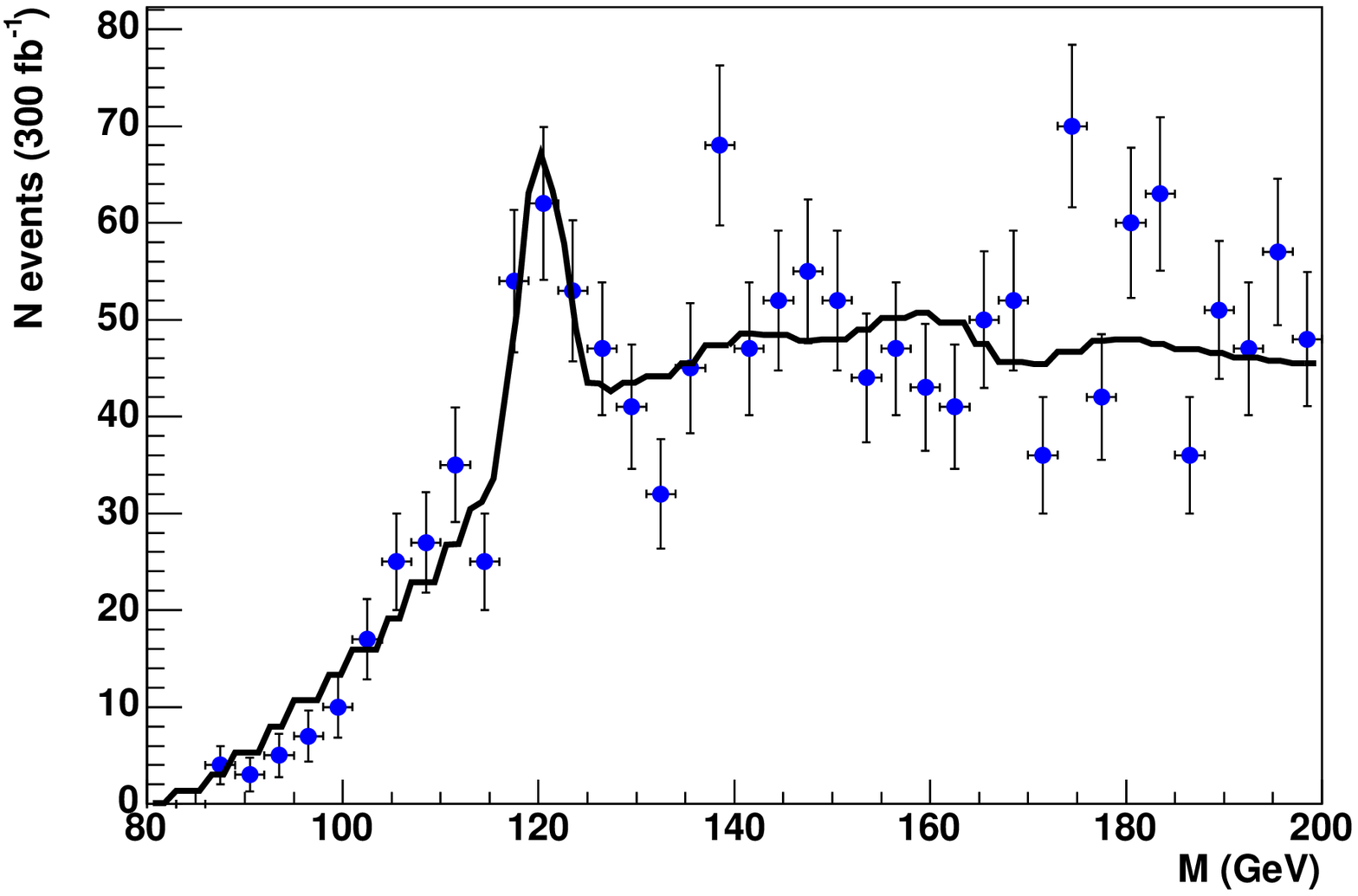}}
\quad
       
\subfigure[]{\includegraphics[width=.5\textwidth]{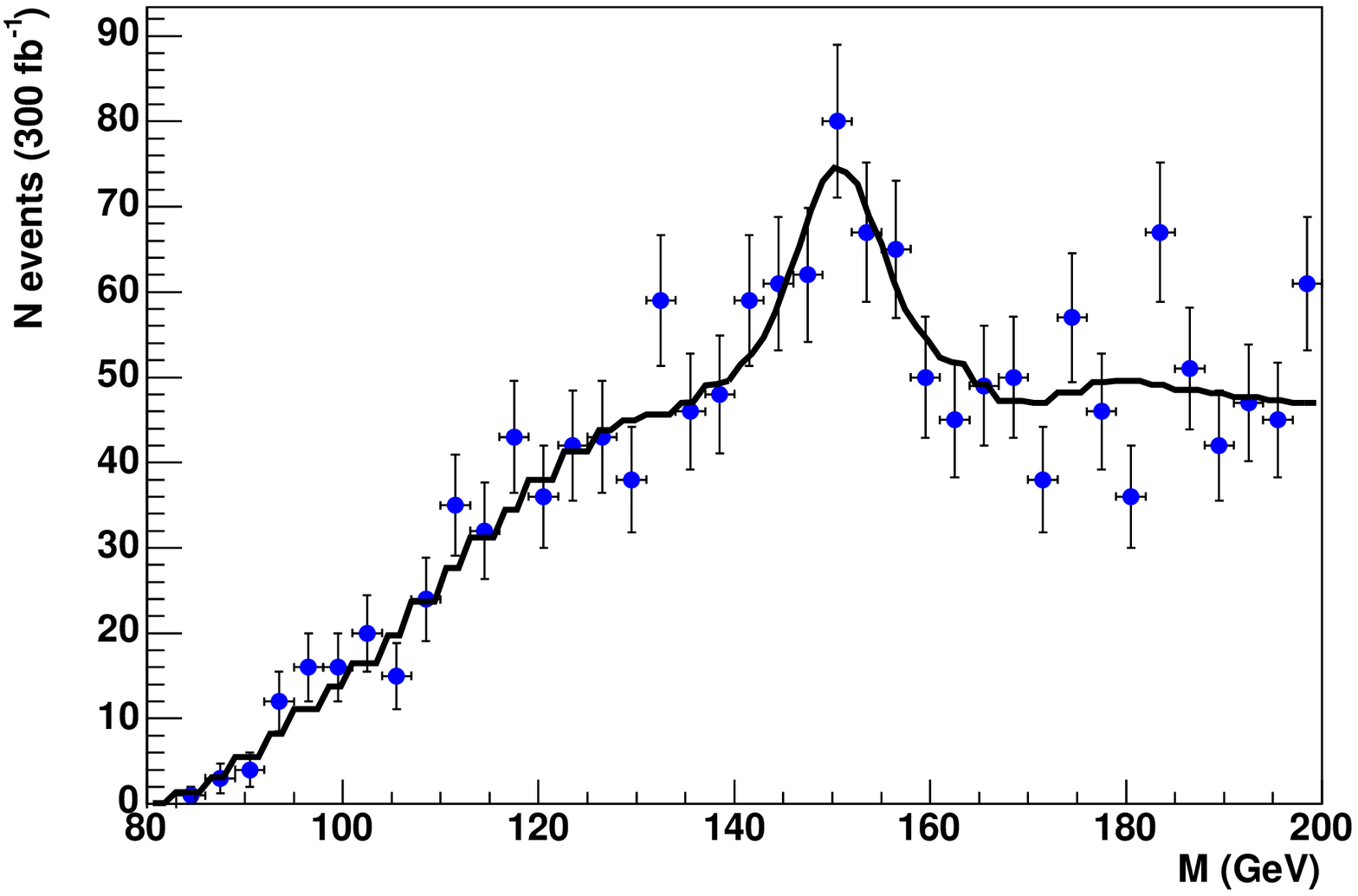}}
        }
\caption{Expected mass distributions given 300~fb$^{-1}$ of data,
collected at 10$^{34}$~cm$^{-2}$~s$^{-1}$ using a R10$+$MU6 trigger for 
the following parameter choices (a) $m_H=120$~GeV and $c_H=0.2$, 
significance is 13.7$\sigma$ (b) $m_H=150$~GeV and $c_H=0.2$, 
significance is 12.7$\sigma$.  (c) $m_H=120$~GeV and $c_H=0.5$, 
significance is 4.3$\sigma$.  (d) $m_H=150$~GeV and $c_H=0.5$, 
significance is 4.3$\sigma$.
\label{fig:highlumistandard}}
\end{figure}

Figure \ref{fig:lowlumistandard} shows the expected mass distributions
for
$m_H=120,150$~GeV and $c_H=0.2,0.5$ for 60~fb$^{-1}$ of data collected
at
2$\times$10$^{33}$~cm$^{-2}$~s$^{-1}$ using a JR25$+$MU6 trigger.
Example distributions for $c_H=0.8$ are not shown as the calculated
significance is less than 3$\sigma$. Figure \ref{fig:highlumistandard}
shows the same distributions but for 300~fb$^{-1}$ of data collected
at 10$^{34}$~cm$^{-2}$~s$^{-1}$ using the R10$+$MU6 trigger.
A summary of the significance for each parameter choice is presented
in Table \ref{tab:signif} for both luminosity/trigger scenarios. The significance 
is approximately 4$\sigma$ for the $c_H=0.5$ parameter points and 
12$\sigma$ for the $c_H=0.2$ parameter points.

For each parameter point, the information obtained from the fitting procedure
 in each pseudo-experiment 
can be used to obtain an RMS spread
 of Higgs masses. This gives a reasonable estimate of the 
 error on the Higgs mass as measured by the forward proton detectors. 
Table \ref{tab:massres} shows the RMS spread for each (significant) 
parameter choice. The R10 trigger strategy (used at high luminosity)
retains only events with asymmetrically tagged protons, whereas the JR25 
trigger (used at low luminosity) retains both symmetric and asymmetric events.
Thus the high luminosity scenario in Table \ref{tab:massres} results in a 
poorer mass measurement than the low luminosity scenario because 
of the lower fraction of symmetric events in the sample. The final trigger strategy
 at ATLAS/CMS will have to balance the need for observation with the opportunity 
 for more precise measurements.
For all parameter choices the mass measurement can be made to better than 2~GeV; for $c_H=0.2$ the mass measurement is better than 0.3~GeV.

It should be noted that the significances obtained for the $c_H=0.2$ parameter points 
are well in excess of 10$\sigma$ and the high rate triggers assumed
in this analysis are not strictly needed for the measurement.
Indeed, the MU6 trigger alone is capable of retaining enough events
for the analysis - the significance for a 120~GeV Higgs boson is 4.5 
for 60~fb$^{-1}$ collected
at 2$\times$10$^{33}$~cm$^{-2}$~s$^{-1}$.
The disadvantage is that the mass measurement is somewhat degraded, due 
to the reduced number of events. 

The $c_H=0.8$ points are observable if the overlap background can be
additionally rejected. 
There are two possibilities related to improvements in the fast-timing system. 
Firstly, if the time-of-flight resolution of the forward detectors is
improved, then both the vertex resolution and overlap rejection is
improved by the same factor.
Secondly, if the central jets can be timed to an accuracy of 70~ps,
using optimal signal filtering in the ATLAS Liquid Argon Calorimeter
\cite{White:2007ha}, a factor of two rejection on the overlap background
would be gained. If the central timing could be performed at the 
10~ps level, then a factor of 12 rejection would be observed. 
The significance of the $c_H=0.8$ parameter point for a 120~GeV Higgs 
 boson increases to 3.2$\sigma$ at high luminosity if  the overlap 
background is additionally rejected by a factor of five. 
Improved overlap rejection would also increase the significance of the 
other parameter points at high luminosity.

\begin{table}[t]
\centering
\begin{tabular}{|c|c|c|c|}
\hline
$m_H$ (GeV) & $c_H$ & \multicolumn{2}{c|}{Significance ($\sigma$)} \\
& & 60~fb$^{-1}$ & 300~fb$^{-1}$ \\
\hline
120 &$0.2$ & 12.7 & 13.7 \\
&$0.5$ & 4.5 & 4.3 \\
&$0.8$ & 2.6 & 2.5 \\
\hline
150 &$0.2$ & 11.9 & 12.7 \\
&$0.5$ & 3.9 & 4.3 \\
&$0.8$ & 2.0  & 2.1 \\
\hline
\end{tabular}
\caption{Significance of central exclusive Higgs boson measurement for
$m_H=120,150$~GeV and $c_H=0.2,0.5,0.8$.
The significance is obtained after using the cuts, trigger strategies
and fitting procedure outlined in the text.
As the overlap background is luminosity dependent, we present results
for both 60~fb$^{-1}$ of data taken at a peak luminosity of
2$\times$10$^{33}$~cm$^{-2}$~s$^{-1}$ and for 300~fb$^{-1}$ of data
taken at a peak luminosity of 10$^{34}$~cm$^{-2}$~s$^{-1}$.
\label{tab:signif}}
\end{table}%

\begin{table}[t]
\centering
\begin{tabular}{|c|c|c|c|}
\hline
$m_H$ (GeV) & $c_H$ & \multicolumn{2}{c|}{$\sigma_{m_H}$~(GeV)} \\
& & 60~fb$^{-1}$ & 300~fb$^{-1}$ \\
\hline
120 &$0.2$ & 0.2 & 0.3 \\
&$0.5$ & 0.8 & 4.5 \\
\hline
150 &$0.2$ & 0.2 & 0.3 \\
&$0.5$ & 1.8 & 2.4 \\
\hline
\end{tabular}
\caption{The error in the Higgs boson mass measurement, $\sigma_{m_H}$, estimated from the 
RMS spread of mass measurements made in
 the pseudo-experiments, for
$m_H=120,150$~GeV and $c_H=0.2,0.5$.
The 60~fb$^{-1}$ of data (taken at a peak luminosity of
2$\times$10$^{33}$~cm$^{-2}$~s$^{-1}$) uses a JR25 $+$ MU6 trigger 
strategy whereas the 300~fb$^{-1}$ of data 
(taken at a peak luminosity of 10$^{34}$~cm$^{-2}$~s$^{-1}$) uses R10 $+$ MU6.
\label{tab:massres}}
\end{table}%

\section{Doubly charged Higgs bosons}

An important feature of the triplet model is the existence of doubly charged
Higgs bosons. 
Although not the focus of this paper, we comment on the possibility to 
observe these using the forward proton system discussed in section 
\ref{sec:fp420}. 
The process of interest is $pp\rightarrow p \oplus H^{++}H^{--}\oplus p$,
where the central system is produced via photon-photon fusion, 
{\it i.e.} $\gamma\gamma \rightarrow H^{++}H^{--}$. 
As two Higgs bosons are produced, the central system will be of higher 
mass than discussed in previous sections and at least one of the protons
will be tagged by detectors placed at 220m from the IP.

The production cross section could be large \cite{han}. 
It was shown in
\cite{nskp,Albrow:2008pn} that the production cross section
was 0.07~fb for a 200~GeV for singly charged scalar pair
production.
However, the cross section for doubly charged Higgs boson
\footnote{The doubly charged Higgs boson mass in the
Machacek-Georgi model is the same as the $H_5^0$ scalar mass, and therefore,
$H^{++}$ with mass $m_{H^{++}}\sim 200-300$ GeV is possible, as it is
heavier than the lightest neutral Higgs $H_1^0$.}
pair production
is a factor of 16 larger due to the factor of two increase in charge.
Thus the cross section for pair-production of a 200~GeV doubly 
charged Higgs boson increases to 1.1~fb.
It should be noted that the photon fusion  production cross
  section decreases roughly by $\sim1/M^{3}$ \cite{KMRpr},
 and so the pair
   production of a 300~GeV doubly charged scalar would be approximately
0.3~fb.

The forward proton acceptance for central systems in the range 
400~GeV-1~TeV is better than 40\% \cite{Bussey:2006vx}. 
This means that we would expect more than 130 events with both 
outgoing protons measured for the pair production of a 200~GeV 
doubly charged Higgs boson given 300~fb$^{-1}$ of data acquisition. 
This decreases to 40 events if the mass of the doubly charged Higgs 
is 300~GeV. 

The events should be retained using the standard electron/muon 
trigger strategies currently in place at ATLAS/CMS. 
A doubly charged Higgs that is heavier than 200~GeV has several possible
decay modes:
$H^{++}\rightarrow W^+W^+$, $W^+H^+$, $H^+H^+$, $l^+l^+$.
In any case, a high fraction of events will contain at least one 
electron/muon in the final state. 

Therefore, forward proton tagging allows the possibility to  
study doubly charged scalars 
that are lighter than around 300~GeV.
Note that, in exclusive production, the background conditions are more
favourable in comparison to the conventional $pp\to H^{++} H^{--} X$ case
considered {\it e.g.} in \cite{han}.
Also, in principle, the forward proton mode may allow
to measure the $H^{\pm\pm}$ mass more accurately than in the inclusive case.


\section{Conclusions}

Searches for the manifestation of the extended Higgs sector
at the LHC may allow new insight in the nature of electroweak
symmetry breaking. The central exclusive production mechanism
 would provide a very powerful tool
to complement the standard strategies at the LHC for studying Higgs
particles. Here we focus on the production of the
neutral Higgs boson of the triplet model in the forward proton
mode.   
We assume a  model with the tree-level value of
the electroweak $\rho$-parameter consistent with experiment,
$\rho=1$.
Although this model is used as a benchmark model for the triplets,
our results are more general.
An extra contribution from other representations enhances the doublet
Yukawa couplings resulting in a different experimental signature to that of the
SM.
We  show that a factor of two enhancement of the fermion couplings 
due to the higher representations implies a significant 
difference to the doublet case.
Let us emphasize that   
in the case of the current model, 
all the fermion couplings to the Higgs boson, which is responsible for the
fermion masses, increase.
This is in contrast with, for instance, the MSSM,
where couplings of up-type and down-type fermions change
from the Standard Model differently, due to the fact that there are 
only doublets in the model.
It is a common feature of higher Higgs representations that the
doublet couplings are enhanced, which thus indicates
that higher representations are involved.

We present a detailed Monte Carlo analysis of the central
exclusive production of a triplet model Higgs boson  
for a number of parameter choices.
For $c_H \leq 0.5$, we have shown that a light $H_1^0$ Higgs boson (of mass 120-150~GeV) can be observed with a 4$\sigma$ (or better) significance if a fixed rate trigger is used. We find that a fixed rate {\it single jet} trigger is optimal at low luminosities whereas a fixed rate {\it forward proton} trigger (i.e. one proton detected at 220m in conjunction with a central jet) is optimal at high luminosities. 

The expected error in the Higgs mass measurement using forward proton detectors is small. For $c_H=0.2$, we find that the mass of the Higgs boson is measured to better than 0.3~GeV, regardless of the luminosity/trigger scenarios. This is due to the excellent mass resolution of the forward detectors and the large number of events. For $c_H=0.5$, there are less events and so the error in the mass measurement increases. However, regardless of the parameter choice, the mass measurement can always be made to better than 2~GeV if a fixed rate single jet trigger is used to retain events in which both protons are tagged at 420~m from the IP.

\section*{Acknowledgments}
We are grateful to Brian Cox, Albert De Roeck, Jeff Forshaw, John Gunion, 
Alan Martin, Risto Orava, Mark Owen, Misha Ryskin,
 Stefan Soldner-Rembold, 
 and Marek Tasevsky for useful discussions. KH gratefully acknowledges support from the Academy of
Finland (Project No. 115032). This work was funded in the UK by the STFC.


\end{document}